\documentclass[final,3p,times]{elsarticle}
\usepackage[latin9]{inputenc}
\usepackage{bm}
\usepackage{amsmath}
\usepackage{amssymb}
\usepackage{graphicx}
\usepackage{esint}
\usepackage{multirow}
\usepackage[subfigure]{graphfig}
\usepackage{rotating}
\usepackage{verbatim}
\usepackage{setspace}
\makeatletter
\newcommand*{\rom}[1]{\expandafter\@slowromancap\romannumeral #1@}

\usepackage{tabularx}\usepackage{subfigure}\usepackage{subfigure}\usepackage{color}

\usepackage{setspace}

\usepackage{bm}

\journal{}

\makeatother

\color{black}

\begin{document}
\title{A new shock-capturing scheme for stiff detonation waves problems}

\author[ad1]{Xi Deng}

\author[ad1]{Honghui Teng}

\author[ad1]{Bin Xie}

\author[ad1]{Feng Xiao}


\address[ad1]{Department of Mechanical Engineering, Tokyo Institute of Technology, \\
	4259 Nagatsuta Midori-ku, Yokohama, 226-8502, Japan.}


\cortext[cor]{Corresponding author:}

\begin{abstract}
A new approach to prevent spurious behavior caused by conventional shock-capturing schemes when solving stiff detonation waves problems is introduced in the present work. Due to smearing of discontinuous solution by the excessive numerical dissipation, conventional shock-capturing schemes have difficulties to obtain the correct location of detonation front without enough grids resolution. To overcome the excessive numerical errors around discontinuities by traditional discretized schemes used in non-reacting high speed compressible flow, we introduce a new shock-capturing scheme in which besides linear function constructed by MUSCL (Monotone Upstream-centered Schemes for Conservation Law) scheme, a step like THINC (Tangent of Hyperbola for INterface Capturing) function is also employed as another candidate in the reconstruction process. The final reconstruction function is determined by boundary variation diminishing (BVD) algorithm by which numerical dissipation around discontinuities can be reduced significantly. The new resulted shock-capturing scheme is named MUSCL-THINC-BVD. One- and two-dimensional comparative numerical tests about stiff detonation waves problems are conducted with the 5th order WENO (Weighted Essentially Non-Oscillatory) and MUSCL-THINC-BVD scheme respectively, which show MUSCL-THINC-BVD scheme can capture the correct position of detonation waves with improved resolution while WENO scheme, in spite of higher order, produces spurious waves. Compared with other existing methods which involves extra treatments by accepting the smeared out discontinuities profiles, the current method obtain the correct but also sharp detonation front by fundamentally reducing numerical dissipation errors from shock-capturing schemes. Thus the proposed approach is an effective but simple method to solve stiff detonation problems. 
\end{abstract}

\maketitle

\section{Introduction}
Standard shock-capturing schemes which resolve discontinuous solutions without numerical oscillations by introducing extra dissipative errors have achieved certain success by far when dealing with non-reacting inviscid compressible flow (Euler equations). However, problems may arise when applying shock-capturing scheme to the reactive Euler equations system which is a non-homogeneous system with source terms to account the reaction effect. When time scales of the chemical reactions are significantly shorter than the hydrodynamic time scales, the problems of numerical stiffness will appear. When dealing with such stiff hyperbolic system, standard shock-capturing schemes with insufficient grids resolution may produce incorrect propagation speed of discontinuities and non-physical spurious waves even with enough temporal resolution. Due to numerical dissipations around discontinuities in shock-capturing scheme, chemical reaction may be triggered too early in the adjacent cell of the discontinuity if numerical smearing out temperature profile contains the value above the ignition temperature. If the reaction is fast, the early triggered chemical reaction will shift the discontinuity thus produce non-physical spurious waves. This phenomenon was first reported in \cite{Yee6}. Hence, to solve the stiff hyperbolic system, a sufficient spatial resolution is as important as temporal one.

During last two decades, spurious phenomenon in simulating stiff hyperbolic system has been investigated by a lot of research work \cite{Nguyen,wang1,Random2000,MinMax,Wang2012,Yee2013}. Various strategies have been proposed to obtain the correct wave propagation speed. To prevent discontinuity front from smearing out by numerical dissipation, ghost fluid/level set and front tracking methods were applied to track the wave front \cite{Nguyen,wang1,wang24}. Another method to reduce numerical dissipation is to locally resolve the detonation wave with a large number of adaptively placed grid cells \cite{wang7}. However, the complexity of scheme design makes the multidimensional calculations less promising. 

Instead of reducing numerical dissipation around discontinuous solution, another kind of strategy is to revise the reaction step by accepting the smeared out discontinuous profile. Considering the spurious phenomenon is caused by earlier triggered chemical reaction by smeared temperature profile, a temperature extrapolation method which constructs a first or second extrapolation from a few grid cells ahead of the shock to obtain the temperature is proposed in \cite{wang14}. However finding the temperature in front of the wave is not trivial especially in higher spatial dimensions \cite{Nguyen}. Instead of extrapolating temperature from the cells in front of shock, the method in which a projection is performed to make the ignition temperature random during reaction step is proposed in \cite{Random2000,Random01,Random02}. Although their methods get success in both one and two dimensional tests, assumption of a priori stiff source term prevents its use for non-stiff problems. To deal with both stiff and non-stiff problem, a fractional-step algorithm termed the MinMax scheme is proposed in \cite{MinMax}. Based on two-value variable reconstruction in each cell, appropriate maximum and minimum values of the unknown are considered to deal with the under-resolved calculations for stiff source conditions. However, only one-dimensional tests are conducted in their work. In the recent work \cite{Wang2012}, high order shock-capturing WENO (Weighted Essentially Non-Oscillatory) is used to update the convection step. During the reaction step, firstly the transition diffused points produced by convection step are identified then the flow variables at these points are extrapolated by a reconstructed polynomial using the idea of subcell resolution method in \cite{Harten}. Although this method can capture the correct location, not only extra steps like solving shock location is necessary, which may has no solution or more than one solution, but also the detonation front is still diffused since existing shock-capturing schemes in spite of high order introduce excessive numerical diffusion errors around discontinuities.

The spurious behavior of high order shock-capturing scheme has been studied recently in \cite{Yee2013}, which reveals that the degree of wrong propagation speed of discontinuities is highly dependent on the manner of shock-capturing schemes in spreading of discontinuities. The fundamental reason behind the spurious phenomenon is that shock-capturing schemes introduce excessive numerical dissipation errors around discontinuities. Thus in our present work, a new shock-capturing scheme which can solve discontinuous solution with substantially reduced numerical dissipation errors will be introduced. During the reconstruction process, besides linear function constructed by MUSCL (Monotone Upstream-centered Schemes for Conservation Law) scheme \cite{Van_Leer}, a step like THINC (Tangent of Hyperbola for INterface Capturing) function \cite{xiao_thinc2} is also employed as another candidate. The final reconstruction function is determined by boundary variation diminishing (BVD) algorithm \cite{Sun, Xie, dengM, dengAIAA} which choose a reconstruction function between MUSCL and THINC by minimizing the variations (jumps) of the reconstructed variables at cell boundaries, which in turn effectively removes the numerical dissipations in numerical solutions. The resulted scheme is named MUSCL-THINC-BVD. 

One- and two-dimensional numerical tests about stiff detonation waves problems show MUSCL-THINC-BVD scheme can capture the sharp and correct detonation front without spurious waves. Compared with existing methods to deal with stiff detonation waves, the MUSCL-THINC-BVD has at least following advantages. 1) Discontinuous solution can be solved sharply without tracking detonation front explicitly like \cite{Nguyen}. 2) Extra extrapolation from neighbor cells like \cite{wang14, Wang2012} is not necessary due to sharp resolved temperature profile. 3) Without extra assumptions, both stiff and non-stiff problems can be solved. 4) Extension to high order version and multi-dimension is straightforward. Actually, these advantages are straightforward since the MUSCL-THINC-BVD scheme reduces the numerical dissipation in shock-capturing schemes fundamentally.     
   
This paper is outlined as follows. In Section~\ref{sec:methods},  after a brief introduction of governing equations and review of the finite volume method in wave-propagation form, the details of the new MUSCL-THINC-BVD scheme for spatial reconstruction are presented. 
In Section~\ref{sec:results}, numerical results of benchmark tests are presented in comparison with high-order schemes. Some concluding remarks end the paper in Section~\ref{sec:conclusion}.

\section{Numerical methods \label{sec:methods}}
We introduce the numerical method in one dimension for the sake of simplicity. Our numerical method can be extended to the multidimensions on structured grids directly in dimension-wise reconstruction fashion. After introduction of control equations and review of the finite volume method in the wave propagation form  \cite{wave1} used in this work, the details about the new MUSCL-THINC-BVD reconstruction scheme will be presented.

\subsection{Reactive Euler equations}
The Euler equations which describe the time-dependent flow of an inviscid compressible and reactive gas with only two chemical states in one space dimension can be written as following form
\begin{equation}
\label{eq:model-nd}
\frac{\partial \textbf{q}}{\partial t} + \frac{\partial f(\textbf{q})}{\partial x} = \phi(\textbf{q}),
\end{equation}
where the vectors of physical variables $\textbf{q}$, flux functions $f$ and source terms $\phi$ are
\begin{equation}
\begin{aligned}
\textbf{q} & = \left ( \rho, \rho u, E, \rho\alpha \right )^{T}, \\
         f & = \left ( \rho u, \rho uu + p,Eu + pu, \rho u\alpha  \right )^{T}, \\
      \phi & = \left ( 0, 0,0, -K(T)\rho\alpha  \right )^{T}.
\end{aligned}
\end{equation}
respectively. The dependent variables $\rho$, $u$, $E$ and $\alpha$ are the density, velocity component in $x$ direction , total energy and mass fraction of unreacted gas, respectively. $p$ is the pressure, $T$ the temperature and $K$ the chemical reaction rate. The pressure is given by
\begin{equation}
p=(\gamma-1)(E-\frac{1}{2}\rho u^2-q_{0}\rho\alpha),
\end{equation}
where $q_{0}$ denotes chemical heat release and $\gamma$ is the ratio of specific heats. The temperature is calculated by
\begin{equation}
T=\dfrac{p}{\rho}.
\end{equation}
For reactive Euler equations, the reaction rate can be modeled with Arrhenius kinetics \cite{random8} by the form
\begin{equation}
K(T)=K_{0}e^{-\frac{T_{ign}}{T}},
\end{equation}
where $K_{0}$ is the reaction rate constant and $T_{ign}$ is the ignition temperature. The reaction rate may also by replaced by Heaviside kinetics with
\begin{equation}
K(T)=-\dfrac{1}{\xi}H(T-T_{ign}),
\end{equation}
where $H(x)=1$ for $x \geq 0$, and $H(x)=0$ for $x < 0$. $\xi$ represents the reaction time. Generally, the stiffness issue becomes more severe with the Heaviside kinetics.

\subsection{Finite volume wave propagation method}
We divide the computational domain into $N$ non-overlapping cell elements, ${\mathcal C}_{i}: x \in [x_{i-1/2},x_{i+1/2} ]$, $i=1,2,\ldots,N$, with a uniform grid with the spacing $\Delta x=x_{i+1/2}-x_{i-1/2}$. For a standard finite volume method, the volume-integrated average value $\bar{\textbf{q}}_{i}(t)$ in the cell $C_{i}$ is defined as
\begin{equation}
\bar{\textbf{q}}_{i}(t) 
\approx \frac{1}{\Delta x} \int_{x_{i-1/2}}^{x_{i+1/2}}
\textbf{q}(x,t) \; dx.
\end{equation}  
Denoting the spatial discretization operator for convection terms in \eqref{eq:model-nd} by  $\mathcal{L}(\bar{\textbf{q}}(t))$,  
the semi-discrete version of the finite volume formulation can be expressed as a system of ordinary differential equations (ODEs)
\begin{equation}
\label{eq:semi-discrete-1d-eq}
\frac{\partial \bar{\textbf{q}}(t)}{\partial t}  = 
\mathcal{L} \left (\bar{\textbf{q}}(t) \right ),
\end{equation}
In the wave-propagation method, the spatial discretization operator for convection terms in cell $C_{i}$ is computed by 
\begin{equation}
\label{eq:spatial-discrete-1d}
\mathcal{L}\left( \bar{\textbf{q}}_{i}(t) \right ) =
-\frac{1}{\Delta x} \left ( 
\mathcal{A}^{+} \Delta \textbf{q}_{i-1/2} + 
\mathcal{A}^{-} \Delta \textbf{q}_{i+1/2} +
\mathcal{A} \Delta \textbf{q}_{i} \right )
\end{equation}	
where  $\mathcal{A}^{+} \Delta \textbf{q}_{i-1/2}$ and
$\mathcal{A}^{-} \Delta \textbf{q}_{i+1/2}$,
are the right- and left-moving fluctuations,
respectively, which enter into the grid cell, and
$\mathcal{A} \Delta \textbf{q}_{i}$ is
the total fluctuation within $C_{i}$. We need to solve Riemann problems to determine these fluctuations. The right- and left-moving fluctuations can be calculated by
\begin{equation}
\mathcal{A}^{\pm} \Delta \textbf{q}_{i-1/2} =
\sum_{k = 1}^{3}
\left [s^{k} 
\left( \textbf{q}_{i-1/2}^{L},\textbf{q}_{i-1/2}^{R} \right )
\right]^{\pm}
\mathcal{W}^{k} \left( \textbf{q}_{i-1/2}^{L},\textbf{q}_{i-1/2}^{R} \right ),
\end{equation} 
where moving speeds $s^{k}$ and the jumps $\mathcal{W}^{k}$ ($k=1,2,3$) of three propagating discontinuities can be solved by Riemann solvers \cite{Riemann} given with the reconstructed values $\textbf{q}_{i-1/2}^{L}$ and $\textbf{q}_{i-1/2}^{R}$ which are computed from the reconstruction functions $\tilde{\textbf{q}}_{i-1}(x)$ and $\tilde{\textbf{q}}_{i}(x)$ to the left and right sides of cell edge $x_{i-1/2}$, respectively. In our simulation, the HLLC Riemann solver \cite{Riemann} will be employed. Similarly, the total fluctuation can be determined by 
\begin{equation}
\label{eq:tfluct-1d}
\mathcal{A} \Delta \textbf{q}_{i} =
\sum_{k = 1}^{3}
\left [s^{k} 
\left( \textbf{q}_{i-1/2}^{R},\textbf{q}_{i+1/2}^{L} \right )
\right]^{\pm}
\mathcal{W}^{k} \left( \textbf{q}_{i-1/2}^{R},\textbf{q}_{i+1/2}^{L} \right )
\end{equation}
We will describe with details about the reconstructions to get these values, $\textbf{q}_{i-1/2}^{L}$ and $\textbf{q}_{i-1/2}^{R}$, at cell boundaries in next subsection as the core part of this paper. 

Given the spatial discretization, we employ three-stage third-order SSP (Strong Stability-Preserving) Runge-Kutta scheme \cite{ssp}
\begin{equation}
\begin{aligned}
&  \bar{\textbf{q}}^{\ast}  = \bar{\textbf{q}}^{n} + 
\Delta t \mathcal{L} \left( \bar{\textbf{q}}^{n} \right ), \\
&\bar{\textbf{q}}^{\ast \ast}  = \frac{3}{4} \bar{\textbf{q}}^{n} + 
\frac{1}{4} \bar{\textbf{q}}^{\ast} +
\frac{1}{4} \Delta t \mathcal{L} \left ( \bar{\textbf{q}}^{\ast} \right ), \\
&\bar{\textbf{q}}_{A}^{n+1}  = \frac{1}{3} \bar{\textbf{q}}^{n} + \frac{2}{3} \bar{\textbf{q}}^{\ast} +
\frac{2}{3} 
\Delta t \mathcal{L}\left ( \bar{\textbf{q}}^{\ast\ast} \right ),
\end{aligned}
\end{equation}
to solve the time evolution ODEs, where $\bar{\textbf{q}}^{\ast}$ and $\bar{\textbf{q}}^{\ast \ast}$ denote the intermediate values at the sub-steps. The $\bar{\textbf{q}}_{A}^{n+1}$ corresponds to the updated value by convection operator in time $\Delta t$. We denote above process to update convection term during $\Delta t$ as $A(\Delta t)$.

In the reaction step, the explicit Euler method is employed as the following formulation
\begin{equation}
\bar{\textbf{q}}_{C}^{n+1}=\bar{\textbf{q}}^{n}+\Delta t \mathcal{S} \left (\bar{\textbf{q}^{n}} \right ),
\end{equation} 
in which the $\bar{\textbf{q}}_{C}^{n+1}$ is the updated value at reaction step and $\mathcal{S} \left (\bar{\textbf{q}} \right )$ is the reaction operator. The above reaction step over $\Delta t$ is denoted by $R(\Delta t)$. To get the numerical solution at time step $n+1$, the Strang-splitting in \cite{wang31} is employed as 
\begin{equation}
\bar{\textbf{q}}^{n+1}=A(\dfrac{\Delta t}{2})R(\dfrac{\Delta t}{Nr})\dots R(\dfrac{\Delta t}{Nr}) A(\dfrac{\Delta t}{2}) \bar{\textbf{q}}^{n},
\end{equation}
where the reaction step is split into $Nr$ sub-steps. In our numerical tests, except special pronouncement we set $Nr=1$.

\subsection{MUSCL-THINC-BVD reconstruction}
In the previous subsection, we left the boundary values, $\textbf{q}_{i-1/2}^{L}$ and $\textbf{q}_{i-1/2}^{R}$, to be determined, which are presented in this subsection. We denote any single variable for reconstruction by $q$ , which can be primitive variable, conservative variable or characteristic variable. In present work, reconstruction regarding to primitive variables will be applied.  

The values $q_{i-1/2}^{L}$ and $q_{i+1/2}^{R}$ at cell boundaries are computed from the piecewise reconstruction functions $\tilde{q}_{i}(x)$ in cell $C_i$. In the present work, we designed the  MUSCL-THINC-BVD reconstruction scheme to capture both smooth and nonsmooth solutions. The BVD algorithm makes use of the MUSCL scheme \cite{Van_Leer} and the THINC scheme \cite{xiao_thinc2} as the candidates for spatial reconstruction. 

In the MUSCL scheme, a piecewise linear function is constructed from the volume-integrated average values $\bar{q}_{i}$, which reads
\begin{equation}
\tilde{q}_{i}(x)^{M}=\bar{q}_{i}+\sigma_{i}(x-x_{i})
\end{equation} 
where $x \in [x_{i-1/2},x_{i+1/2}]$ and $\sigma_{i}$ is the slope defined at the cell center $x_{i}=\frac{1}{2}(x_{i-1/2}+x_{i+1/2})$. To prevent numerical oscillation, a slope limiter \cite{Van_Leer,finite} is used to get numerical solutions satisfying the TVD property. We denote the reconstructed values at cell boundaries from MUSCL reconstruction as $q_{i-1/2}^{R,ML}$ and $q_{i+1/2}^{L,M}$. 
The MUSCL scheme, in spite of popular use in various numerical models, has excessive numerical dissipation around discontinuous solution which tends to produce spurious phenomenon for stiff detonation waves . 

Being another reconstruction candidate, the THINC \cite{xiao_thinc,xiao_thinc2} uses the hyperbolic tangent function, which is a differentiable and monotone function that fits well a step-like discontinuity. The THINC reconstruction function is written as
\begin{equation}
\tilde{q}_{i}(x)^{T}=\bar{q}_{min}+\dfrac{\bar{q}_{max}}{2} \left(1+\theta~\tanh \left(\beta \left(\dfrac{x-x_{i-1/2}}{x_{i+1/2}-x_{i-1/2}}-\tilde{x}_{i}\right)\right)\right),
\end{equation} 
where $\bar{q}_{min}=min(\bar{q}_{i-1},\bar{q}_{i+1})$, $\bar{q}_{max}=max(\bar{q}_{i-1},\bar{q}_{i+1})-\bar{q}_{min}$ and $\theta=sgn(\bar{q}_{i+1}-\bar{q}_{i-1})$. The jump thickness is controlled by parameter $\beta$. In our numerical tests shown later a constant value of $\beta=1.8$ is used. The unknown $\tilde{x}_{i}$, which represents the location of the jump center, is computed from $ \bar{q}_{i} = \frac{1}{\Delta x} \int_{x_{i-1/2}}^{x_{i+1/2}} \tilde{q}_{i}(x)^{THINC} \; dx$. Then the reconstructed values at cell boundaries by THINC function can be expressed by 
\begin{equation}
\label{thniclr}
\begin{aligned}
&q_{i+1/2}^{L,T}=\bar{q}_{min}+\dfrac{\bar{q}_{max}}{2} \left(1+\theta \dfrac{\tanh(\beta)+A}{1+A~\tanh(\beta)}\right)\\
&q_{i-1/2}^{R,T}=\bar{q}_{min}+\dfrac{\bar{q}_{max}}{2} \left(1+\theta~ A\right)
\end{aligned}
\end{equation}
where $A=\frac{B/\cosh(\beta)-1}{\tanh(\beta)}$ and $B=\exp(\theta~\beta(2~C-1))$, where $C=\dfrac{\bar{q}_{i}-\bar{q}_{min}+\epsilon}{\bar{q}_{max}+\epsilon}$ and $\epsilon=10^{-20}$ is a mapping factor to project the physical fields onto $[0,1]$.

The final effective reconstruction function is determined by the BVD algorithm. Several practical versions of BVD algorithm have been proposed recently by \cite{Sun,Xie,dengAIAA,dengM}. In present work the version in \cite{dengM} will be implemented. In BVD algorithm of \cite{dengM}, reconstruction function is chosen between $\tilde{q}_{i}(x)^{M}$ and $\tilde{q}_{i}(x)^{T}$ so that the variations of the reconstructed values at cell boundaries are minimized. The previous studies \cite{Sun,Xie,dengAIAA,dengM} have shown that BVD algorithm prefers the THINC reconstruction $\tilde{q}_{i}(x)^{T}$ within a cell where a discontinuity exists, by which way the numerical dissipation will be reduced significantly around discontinuous solutions. It is sensible that the THINC reconstruction should only be employed when a discontinuity is detected. In practice, a cell where a discontinuity may exist can be identified by the following conditions
\begin{equation}
\label{eq:thincC}
\begin{aligned}
&\delta<C<1-\delta,\\
&(\bar{q}_{i+1}-\bar{q}_{i})(\bar{q}_{i}-\bar{q}_{i-1})>0,
\end{aligned}
\end{equation}
where $\delta$ is a small positive (e.g.,$10^{-4}$). 

The details of the reconstruction function of MUSCL-THINC-BVD scheme in \cite{dengM} reads 
\begin{equation}
\tilde{q}_{i}(x)^{BVD}=\left\{
\begin{array}{l}
\tilde{q}_{i}(x)^{T}~~~\mathrm{if}~\delta<C<1-\delta,~\mathrm{and}~ (\bar{q}_{i+1}-\bar{q}_{i})(\bar{q}_{i}-\bar{q}_{i-1})>0,~\mathrm{and}~TBV_{i,min}^{T}<TBV_{i,min}^{M}\\
\tilde{q}_{i}(x)^{M}~~~\mathrm{otherwise}
\end{array}
\right..
\end{equation}
where the minimum value of total boundary variation (TBV) $TBV_{i,min}^{P}$, for reconstruction function $P =T~or~M$ representing THINC and MUSCL reconstruction respectively, is defined as
\begin{equation}
\begin{aligned}
TBV_{i,min}^{P}=min(&|q_{i-1/2}^{L,M}-q_{i-1/2}^{R,P}|+|q_{i+1/2}^{L,P}-q_{i+1/2}^{R,M}|,|q_{i-1/2}^{L,T}-q_{i-1/2}^{R,P}|+|q_{i+1/2}^{L,P}-q_{i+1/2}^{R,T}|,
\\&|q_{i-1/2}^{L,M}-q_{i-1/2}^{R,P}|+|q_{i+1/2}^{L,P}-q_{i+1/2}^{R,T}| ,|q_{i-1/2}^{L,T}-q_{i-1/2}^{R,P}|+|q_{i+1/2}^{L,P}-q_{i+1/2}^{R,M}|) .
\end{aligned}
\end{equation}

Thus, THINC reconstruction function will be employed in the targeted cell if the minimum TBV value of THINC is smaller than that of MUSCL. As stated in previous work \cite{Sun,dengAIAA,dengM}, the BVD algorithm will realize the polynomial interpolation for smooth solution while for discontinuous solution a step like function will be preferred. 

As shown in numerical tests in this paper, discontinuities including detonation front can be resolved by the MUSCL-THINC-BVD scheme with substantially reduced numerical dissipation in comparison with standard 5th order WENO scheme \cite{Jiang}. It is noted that the MUSCL-THINC-BVD is essentially a shock capturing scheme which can solve discontinuous solution sharply without explicit special treatment about the detonation front like other methods.
 
\section{Numerical results \label{sec:results}}
In this section, comparative tests in one- and two- dimensions are conducted with 5th order WENO scheme\cite{Jiang} and the proposed MUSCL-THINC-BVD scheme.  

\subsection{C-J detonation wave with Arrhenius law}
\begin{figure}
 	\subfigure[WENO]{\centering\includegraphics[scale=0.3,trim={1.0cm 1.0cm 1.0cm 1.0cm},clip]{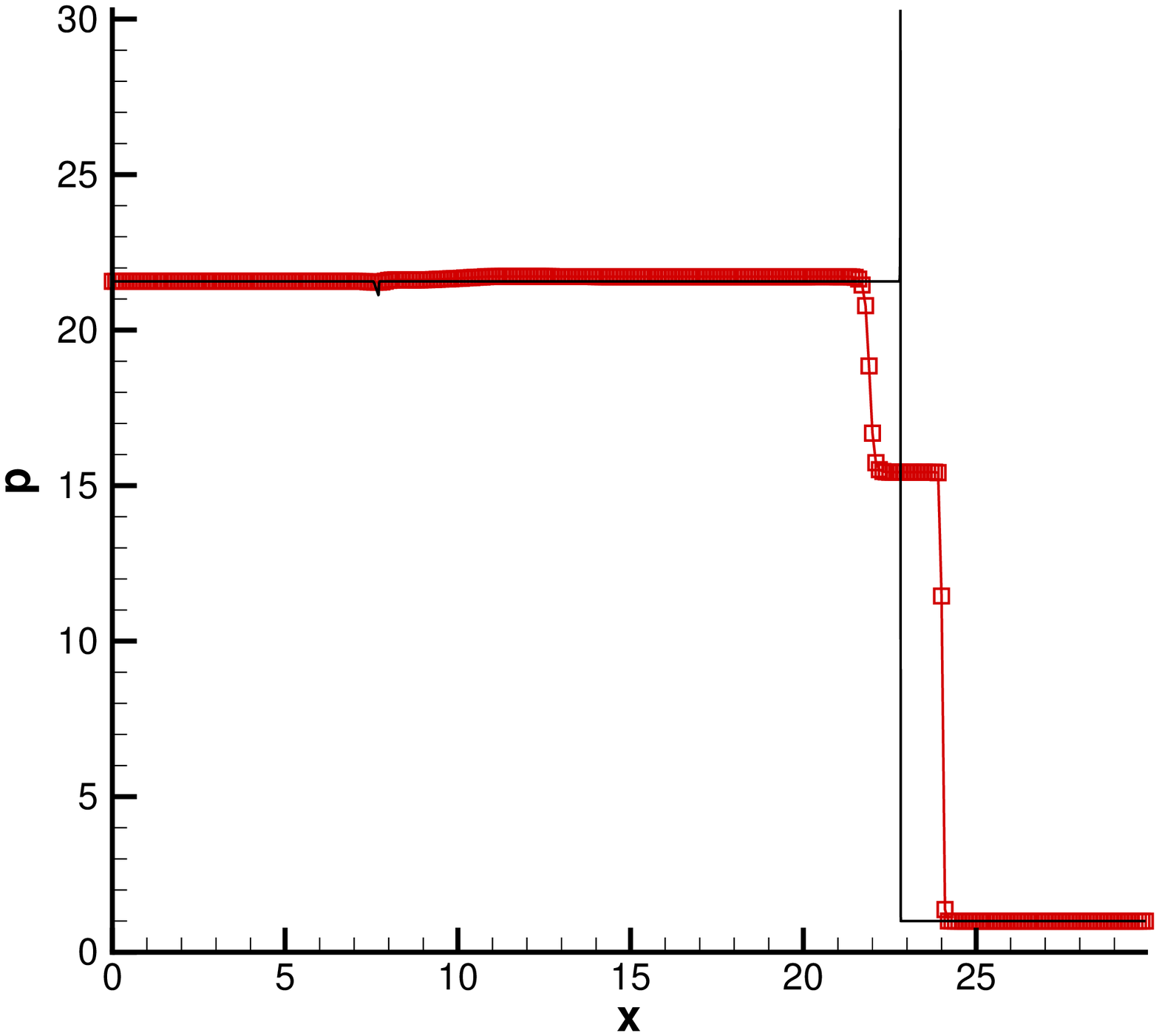}}
 	\subfigure[MUSCL-THINC-BVD]{\centering\includegraphics[scale=0.3,trim={1.0cm 1.0cm 1.0cm 1.0cm},clip]{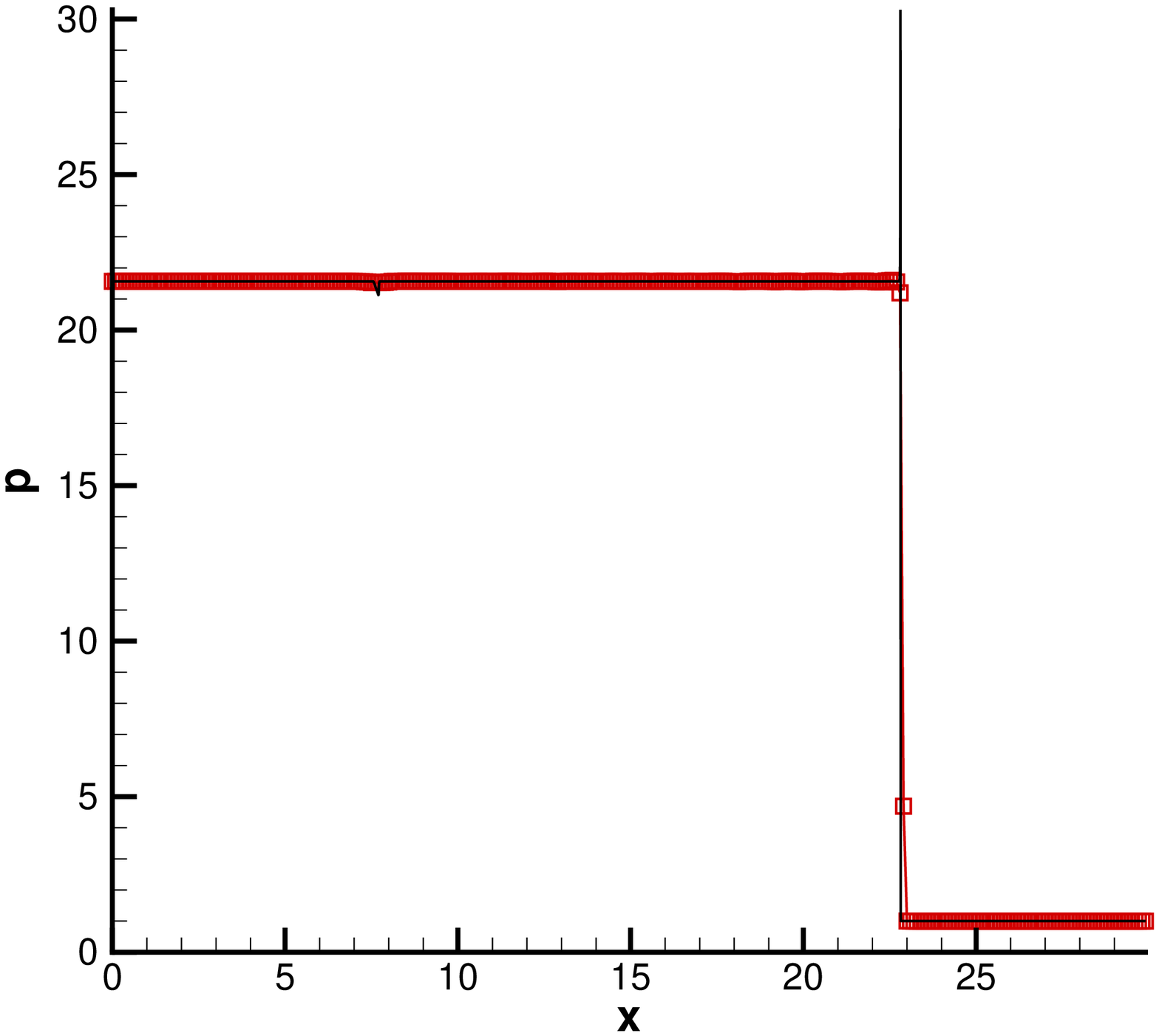}}
 	\protect\caption{Numerical results of pressure field for C-J detonation wave with Arrehenius law. Reference solutions are represented by black solid lines while numerical solutions are represented by red lines with symbols. Comparisons are made between the WENO and MUSCL-THINC-BVD scheme. \label{fig:ex41P}}	
 \end{figure}

\begin{figure}
	\subfigure[WENO]{\centering\includegraphics[scale=0.3,trim={1.0cm 1.0cm 1.0cm 1.0cm},clip]{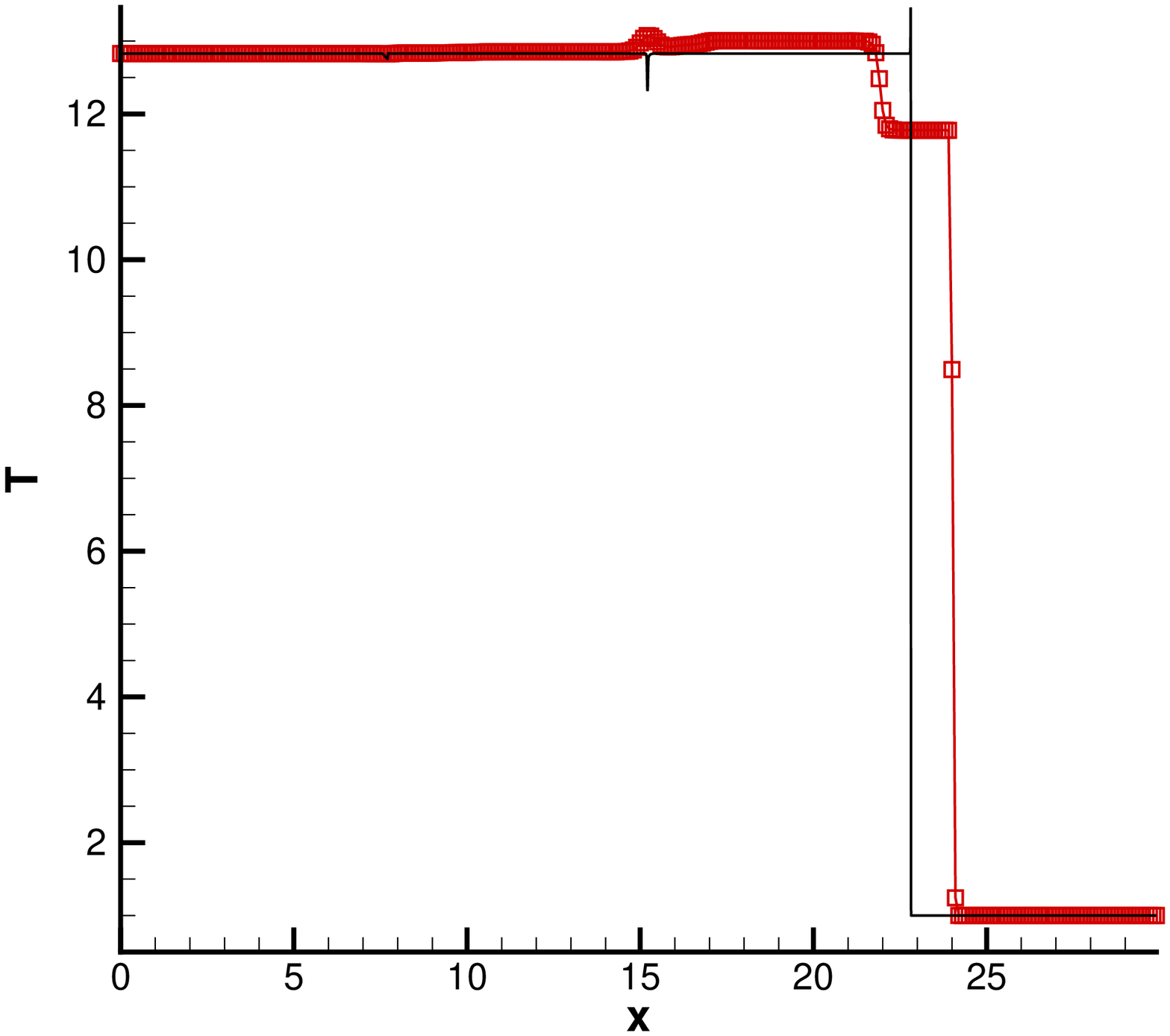}}
	\subfigure[MUSCL-THINC-BVD]{\centering\includegraphics[scale=0.3,trim={1.0cm 1.0cm 1.0cm 1.0cm},clip]{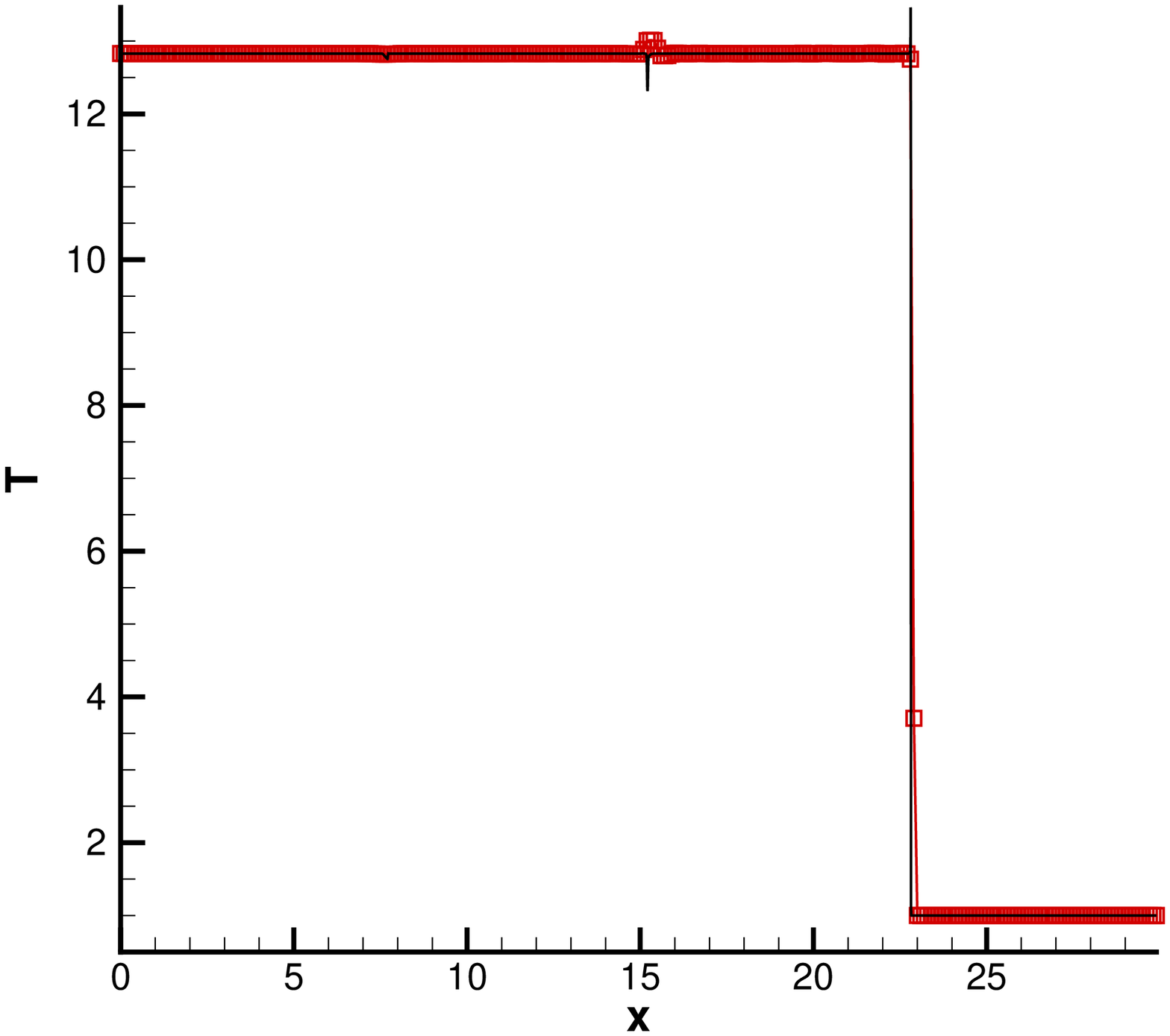}}
	\protect\caption{The same as Fig.~\ref{fig:ex41P} but for temperature field. \label{fig:ex41T}}	
\end{figure}

\begin{figure}
	\subfigure[WENO]{\centering\includegraphics[scale=0.3,trim={1.0cm 1.0cm 1.0cm 1.0cm},clip]{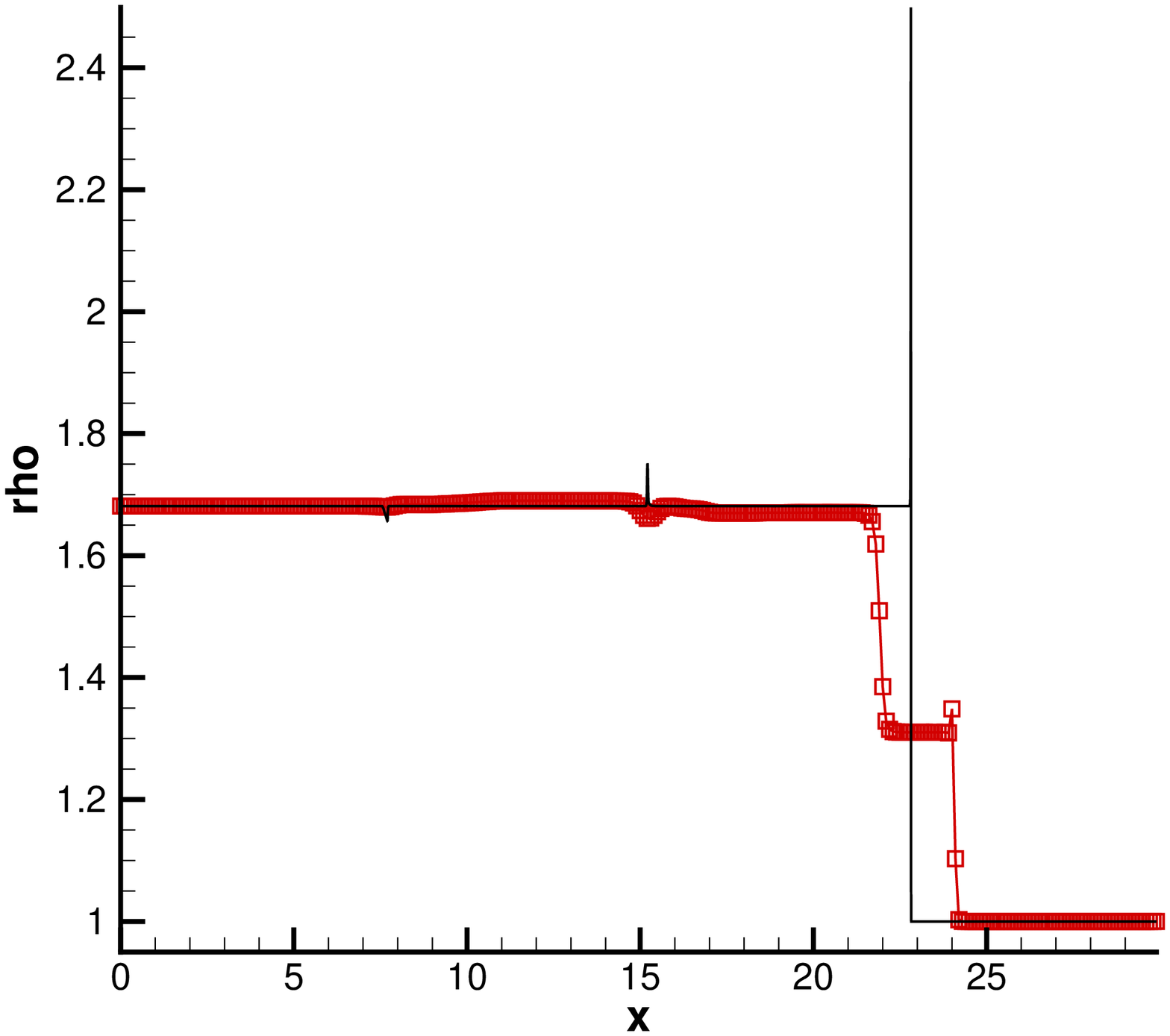}}
	\subfigure[MUSCL-THINC-BVD]{\centering\includegraphics[scale=0.3,trim={1.0cm 1.0cm 1.0cm 1.0cm},clip]{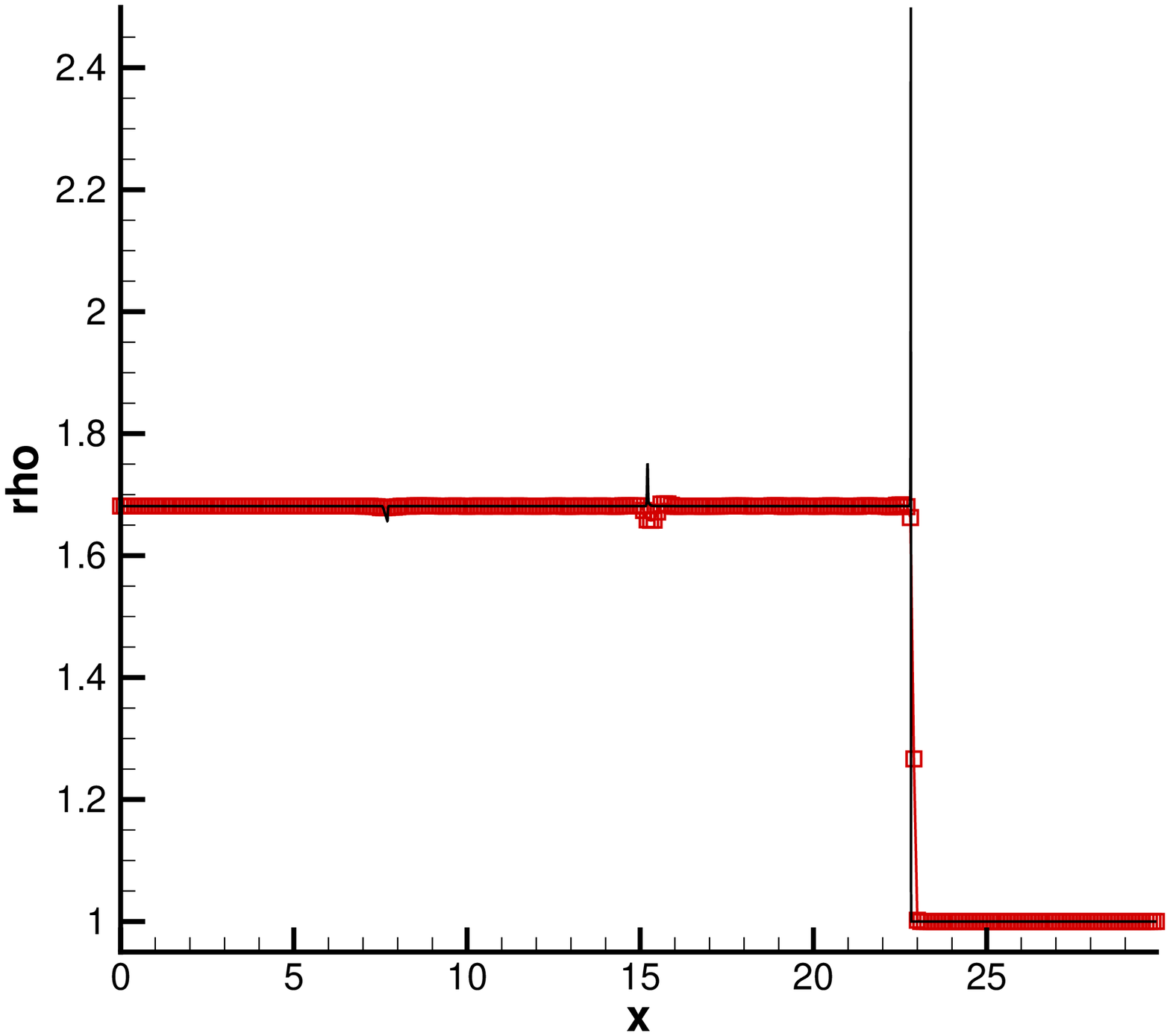}}
	\protect\caption{The same as Fig.~\ref{fig:ex41P} but for density field. \label{fig:ex41rho}}	
\end{figure}

\begin{figure}
	\subfigure[WENO]{\centering\includegraphics[scale=0.3,trim={1.0cm 1.0cm 1.0cm 1.0cm},clip]{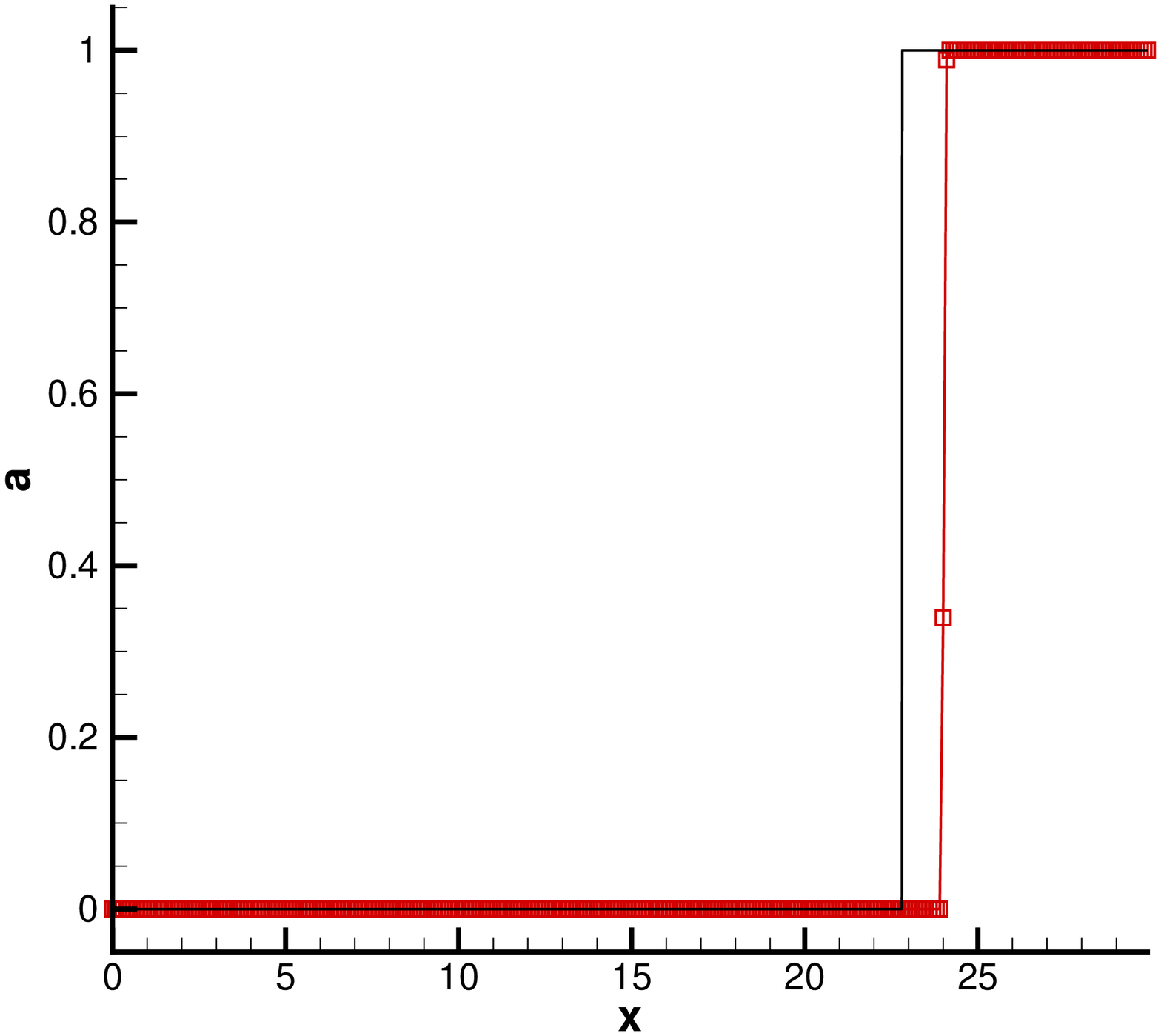}}
	\subfigure[MUSCL-THINC-BVD]{\centering\includegraphics[scale=0.3,trim={1.0cm 1.0cm 1.0cm 1.0cm},clip]{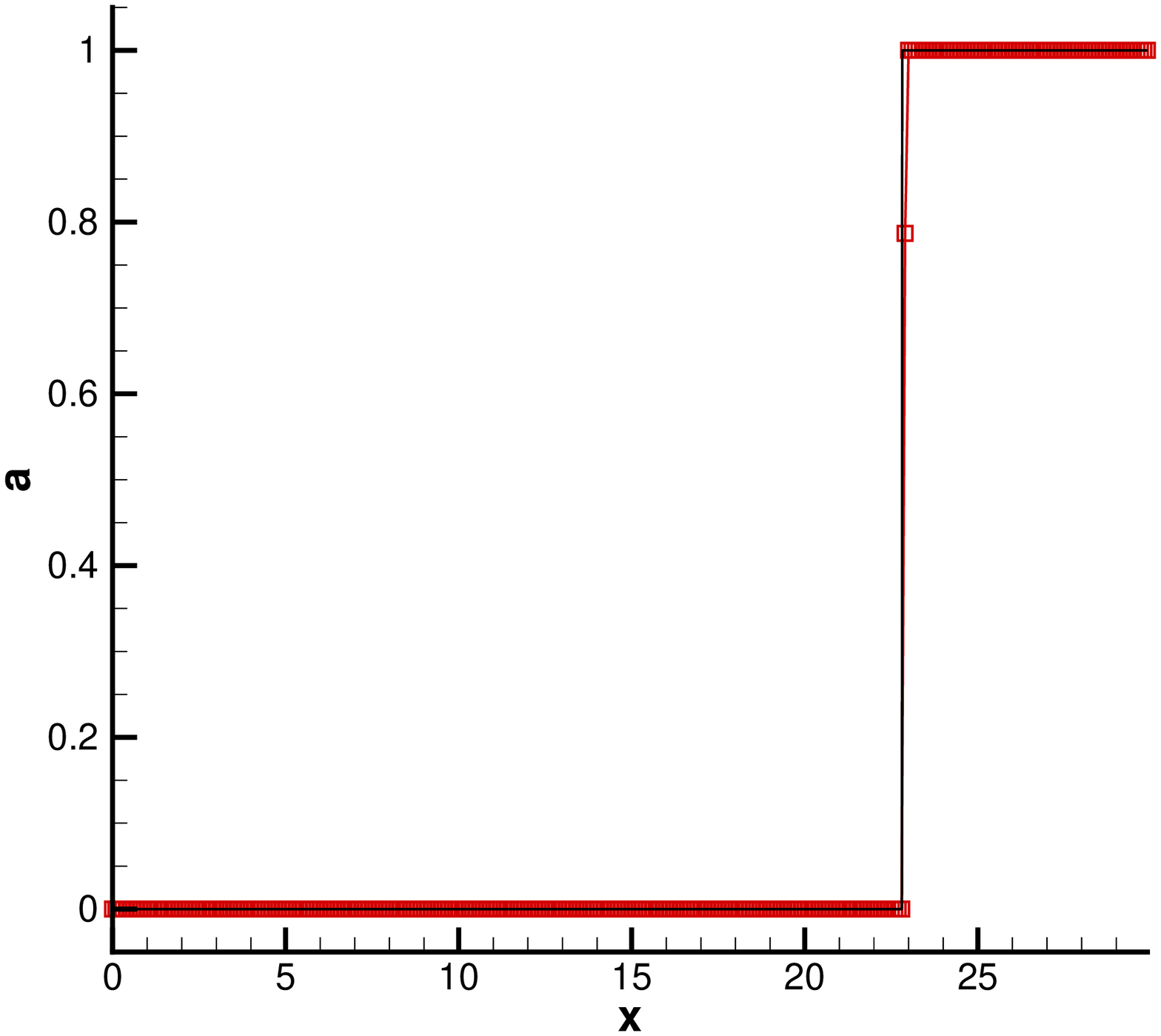}}
	\protect\caption{The same as Fig.~\ref{fig:ex41P} but for mass fraction. \label{fig:ex41a}}	
\end{figure}

\begin{figure}
	\subfigure[WENO]{\centering\includegraphics[scale=0.3,trim={1.0cm 1.0cm 1.0cm 1.0cm},clip]{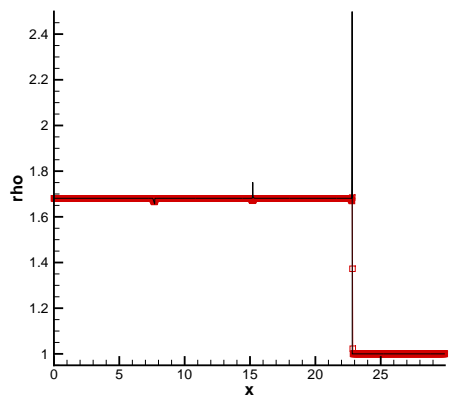}}
	\subfigure[MUSCL-THINC-BVD]{\centering\includegraphics[scale=0.3,trim={1.0cm 1.0cm 1.0cm 1.0cm},clip]{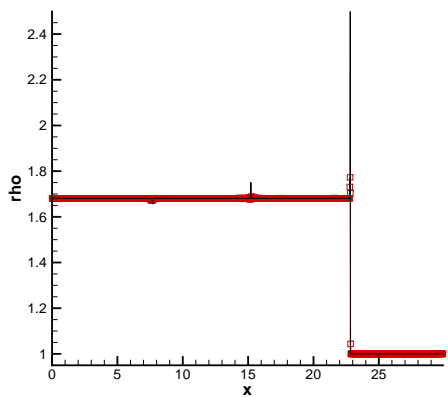}}
	\protect\caption{The same as Fig.~\ref{fig:ex41rho} but with finer mesh. \label{fig:ex41rhoF}}	
\end{figure}

The C-J detonation wave in Chapman-Jouguet model with Arrhenius source term is considered in this case, which has also been studied in \cite{Random2000,MinMax,Wang2012}. Initially, totally burnt gas is set on the left-hand side while totally unburnt gas on the right. We set the ration of specific heats $\gamma=1.4$, the heat release $q_{0}=25$ the ignition temperature $T_{ign}=25$ and $K_{0}=16418$. Given any initial state $(\rho_{0},u_{0},p_{0},1.0)$ of right side, the C-J initial state $(\rho_{CJ},u_{CJ},p_{CJ},0.0)$ on the left side can be obtained by \cite{random5,random8}. The computation domain is $[0,30]$ where initial discontinuity is located at $x=10$. In present simulation, we set $\rho_{0}=1.0$, $u_{0}=0.0$ and $p_{0}=1.0$. The computation is evolved until $t=1.8$. The computation is conducted by standard 5th order WENO and proposed MUSCL-THINC-BVD scheme respectively with cell numbers $N=300$. For WENO scheme, the CFL=0.05 while CFL=0.1 for MUSCL-THINC-BVD scheme. The numerical results regarding to pressure, temperature, density and mass fraction are displayed in Fig.~\ref{fig:ex41P}-\ref{fig:ex41a} in which the reference solution is calculated by standard 5th order WENO scheme with cell numbers $N=10000$. From the results, the correct speed of the C-J detonation can be captured by MUSCL-THINC-BVD scheme with larger CFL number while a spurious weak detonation appears ahead of the detonation wave by WENO scheme. In spite of its higher order, WENO scheme will introduce too much numerical diffusion errors around discontinuities, which causes the spurious numerical solutions. The comparative result regarding to density field between WENO and MUSCL-THINC-BVD scheme is also made with finer grid cells $N=1500$ in Fig.~\ref{fig:ex41rhoF}. It can be seen that the combustion spike can be resolved by the MUSCL-THINC-BVD scheme due to it's less diffusive around critical regions.  

\subsection{C-J detonation wave with the Heaviside model \label{CJ-H}}
\begin{figure}
	\subfigure[WENO]{\centering\includegraphics[scale=0.3,trim={1.0cm 1.0cm 1.0cm 1.0cm},clip]{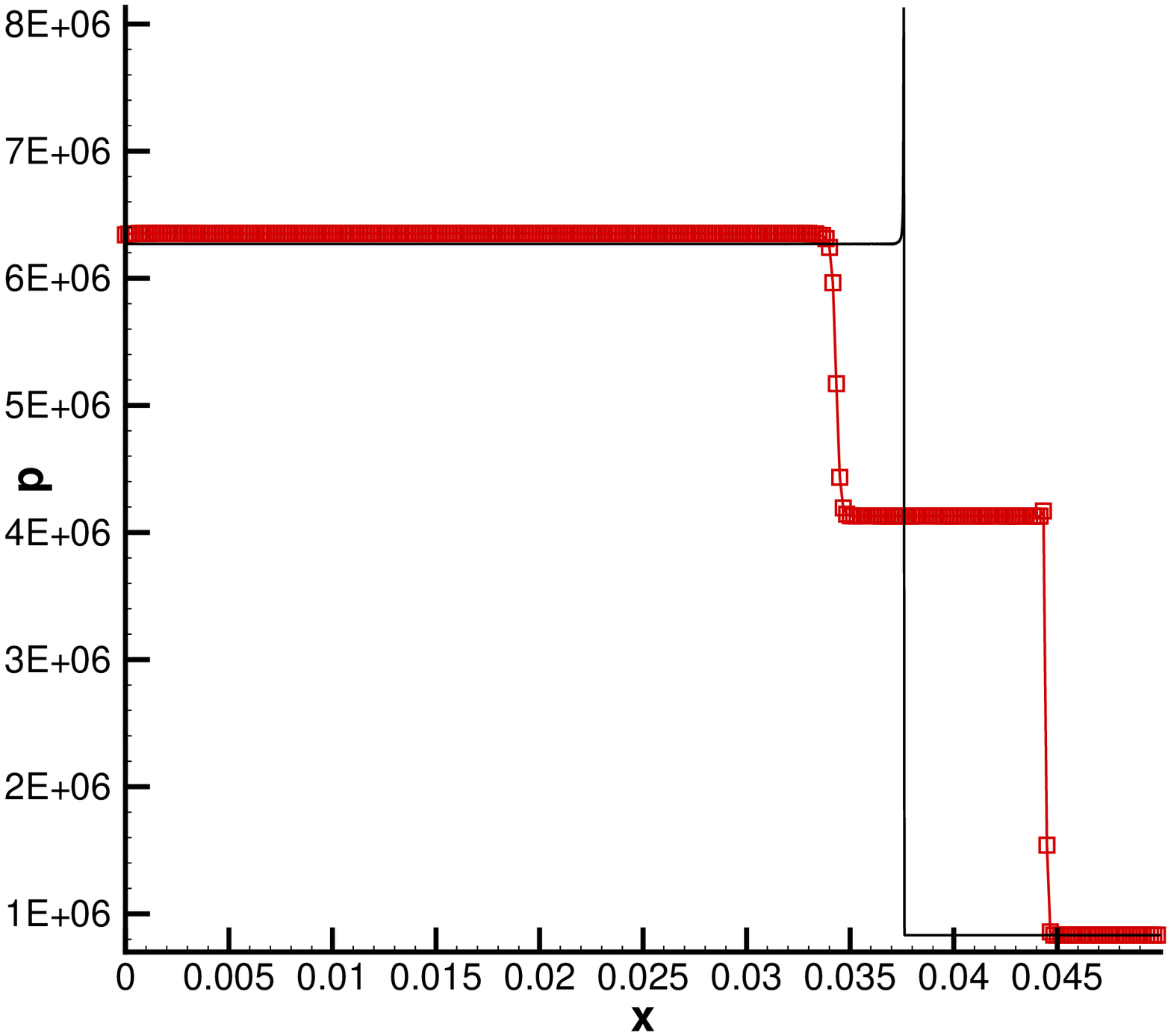}}
	\subfigure[MUSCL-THINC-BVD]{\centering\includegraphics[scale=0.3,trim={1.0cm 1.0cm 1.0cm 1.0cm},clip]{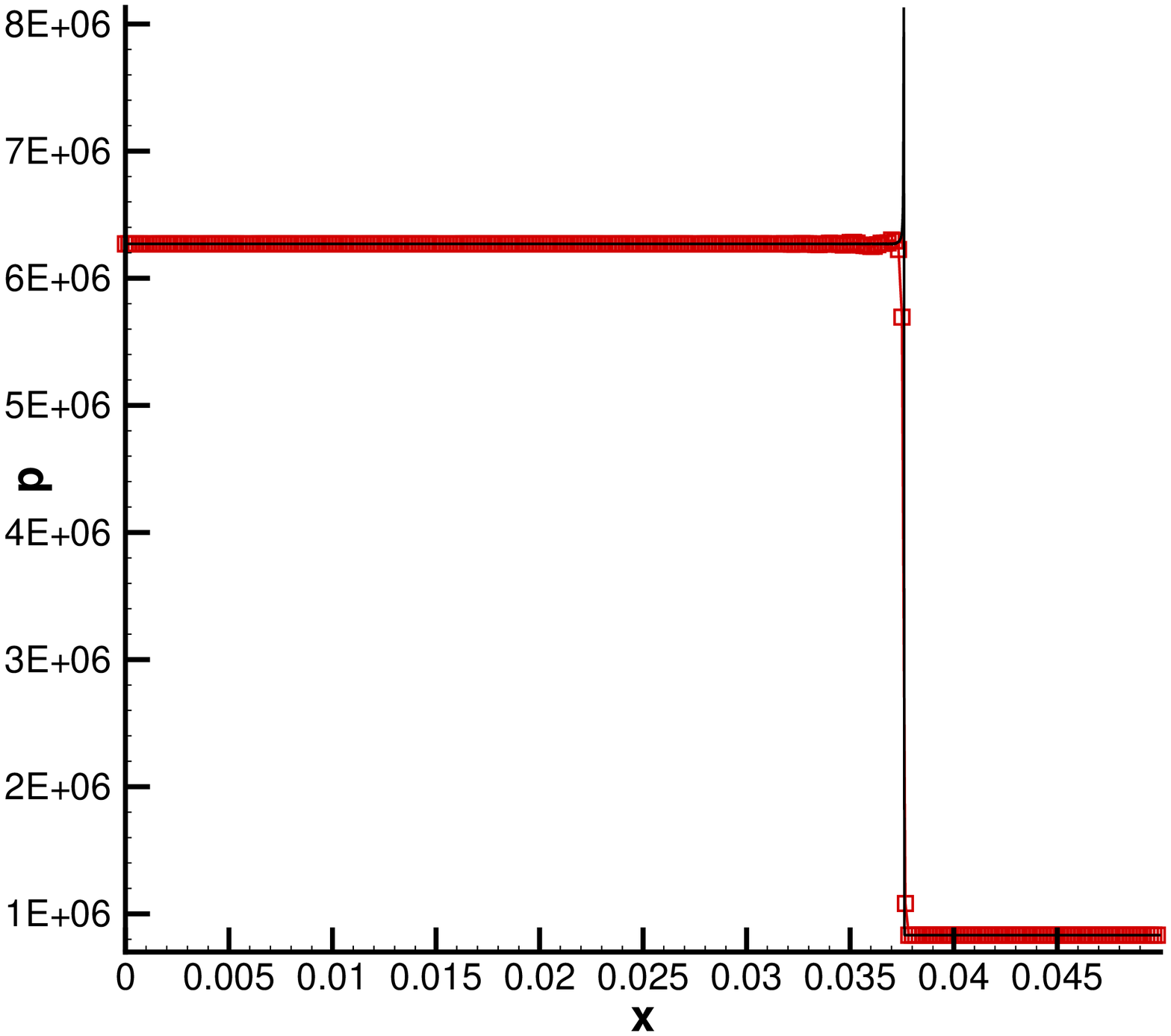}}
	\protect\caption{Numerical results of pressure field for C-J detonation wave with the Heaviside model. Reference solutions are represented by black solid lines while numerical solutions are represented by red lines with symbols. Comparisons are made between the WENO and MUSCL-THINC-BVD scheme. \label{fig:ex42P}}	
\end{figure}

\begin{figure}
	\subfigure[WENO]{\centering\includegraphics[scale=0.3,trim={1.0cm 1.0cm 1.0cm 1.0cm},clip]{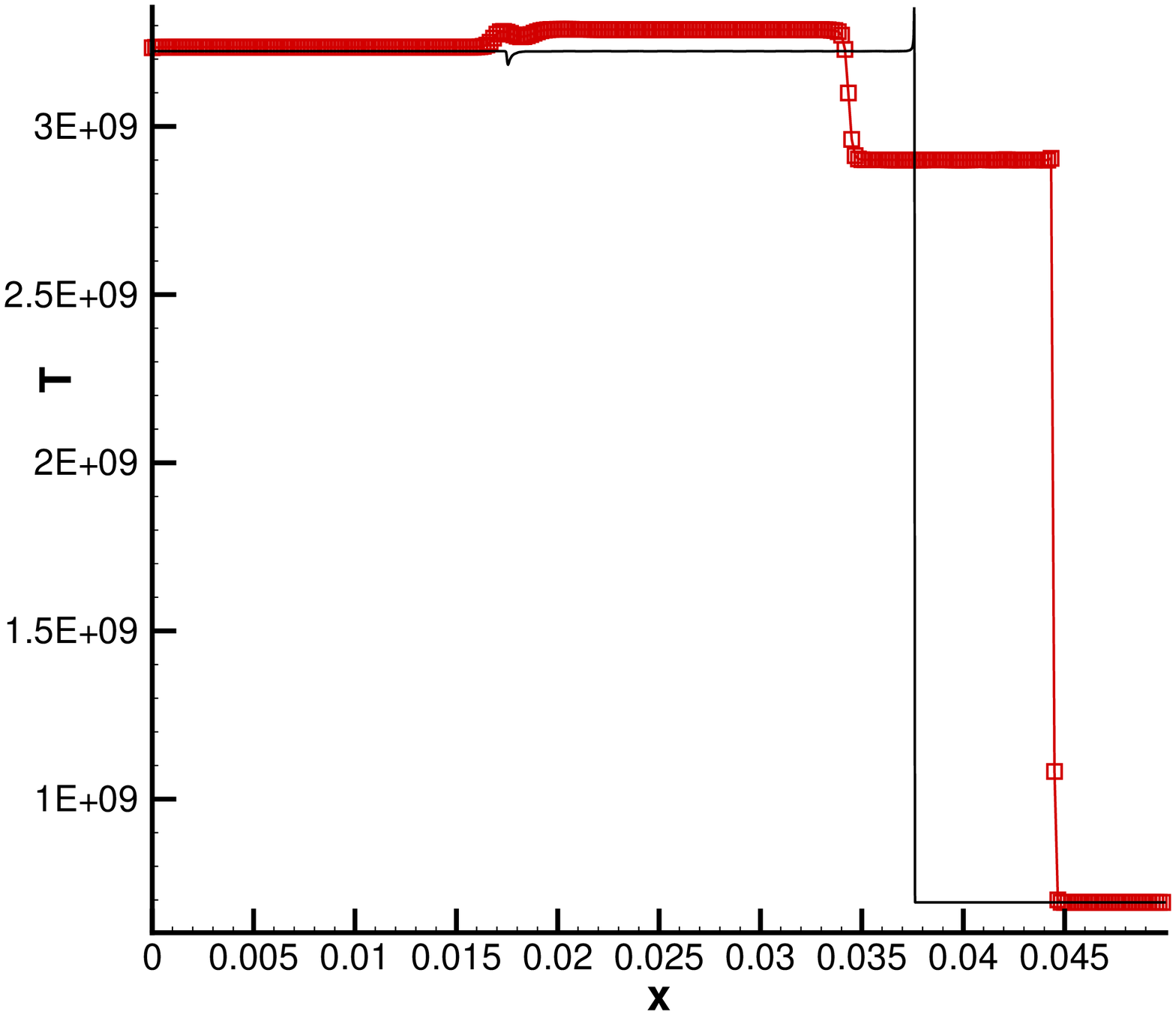}}
	\subfigure[MUSCL-THINC-BVD]{\centering\includegraphics[scale=0.3,trim={1.0cm 1.0cm 1.0cm 1.0cm},clip]{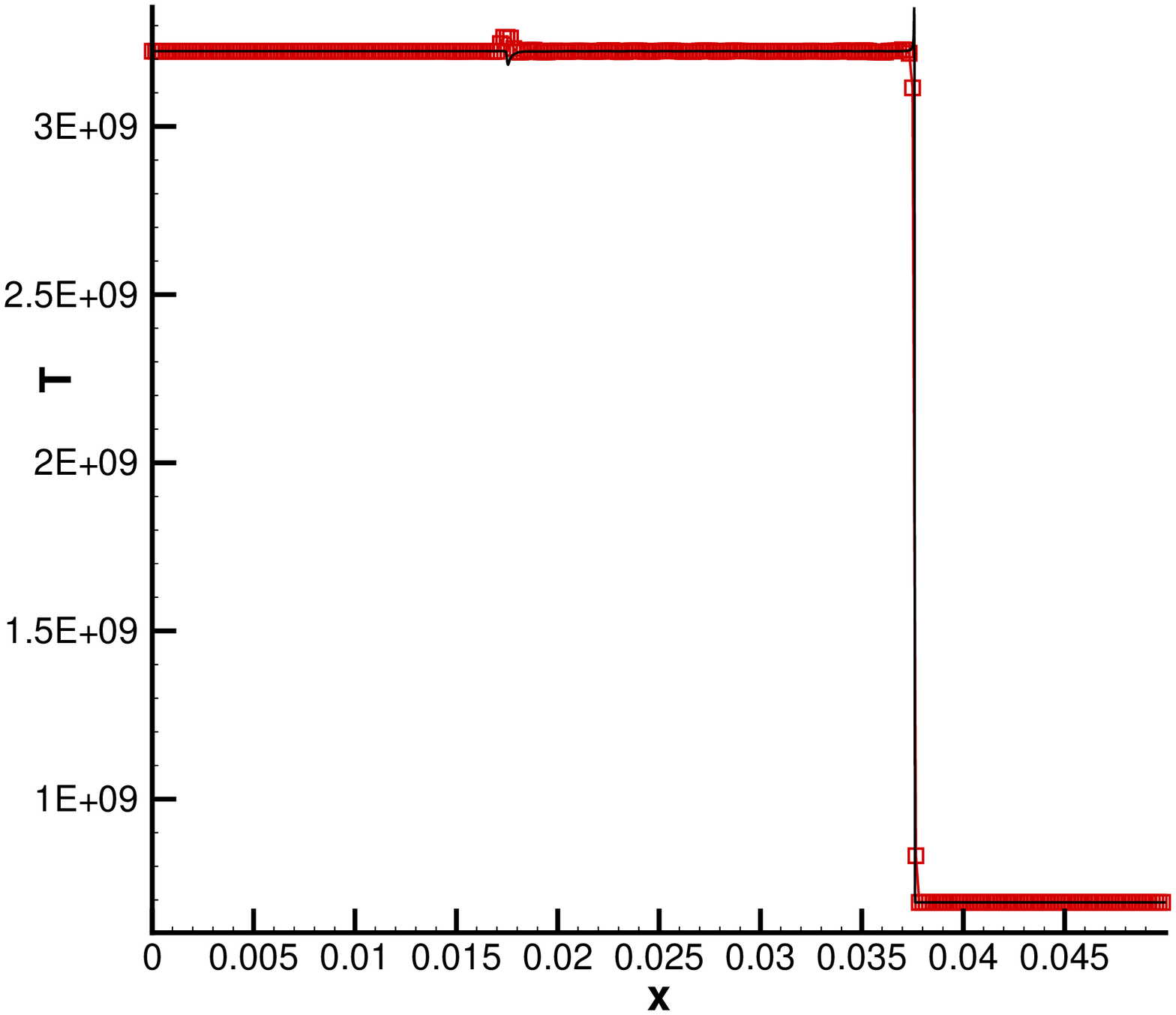}}
	\protect\caption{The same as Fig.~\ref{fig:ex42P} but for temperature field. \label{fig:ex42T}}	
\end{figure}

\begin{figure}
	\subfigure[WENO]{\centering\includegraphics[scale=0.3,trim={0.6cm 1.0cm 1.0cm 1.0cm},clip]{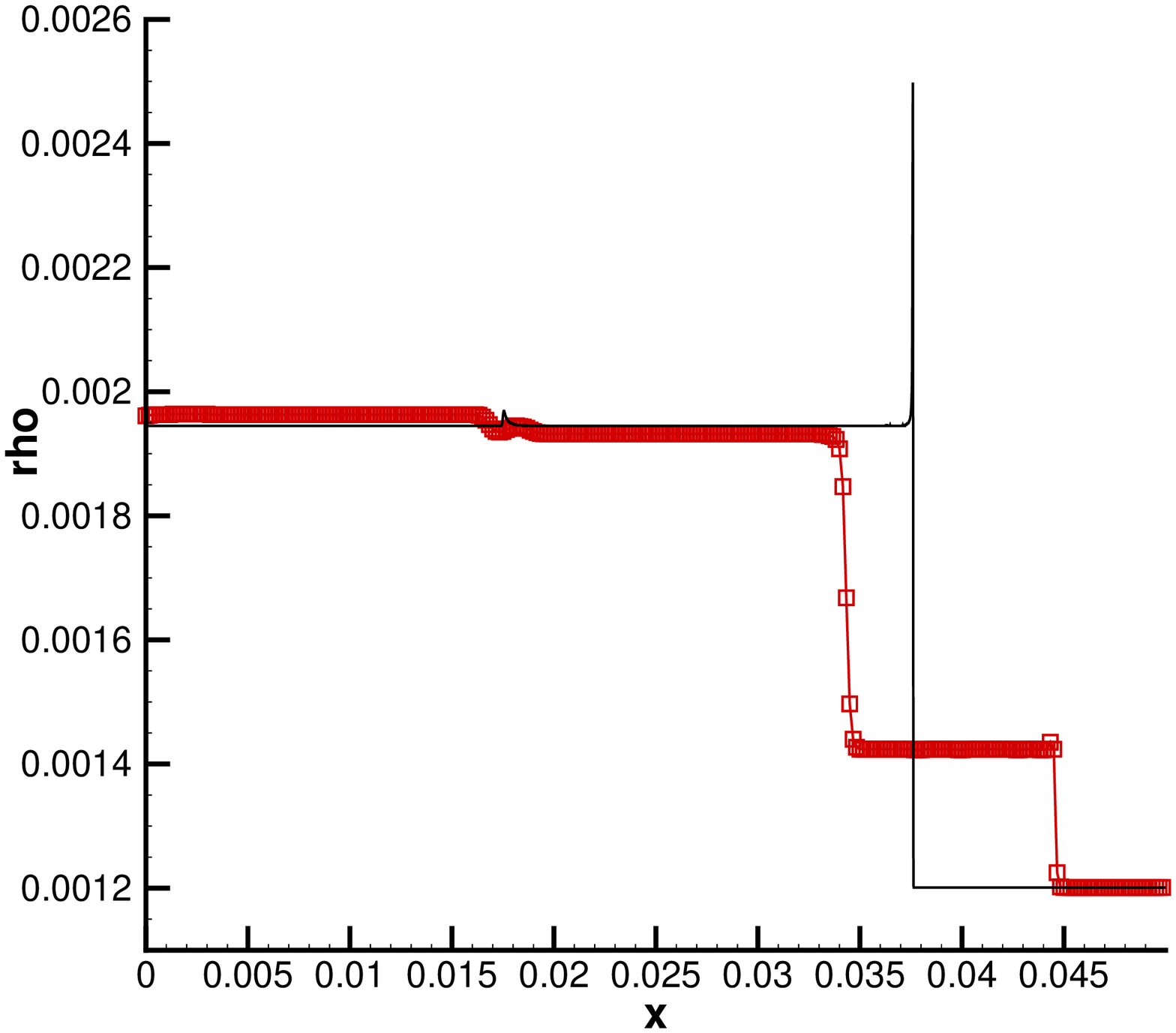}}
	\subfigure[MUSCL-THINC-BVD]{\centering\includegraphics[scale=0.3,trim={0.6cm 1.0cm 1.0cm 1.0cm},clip]{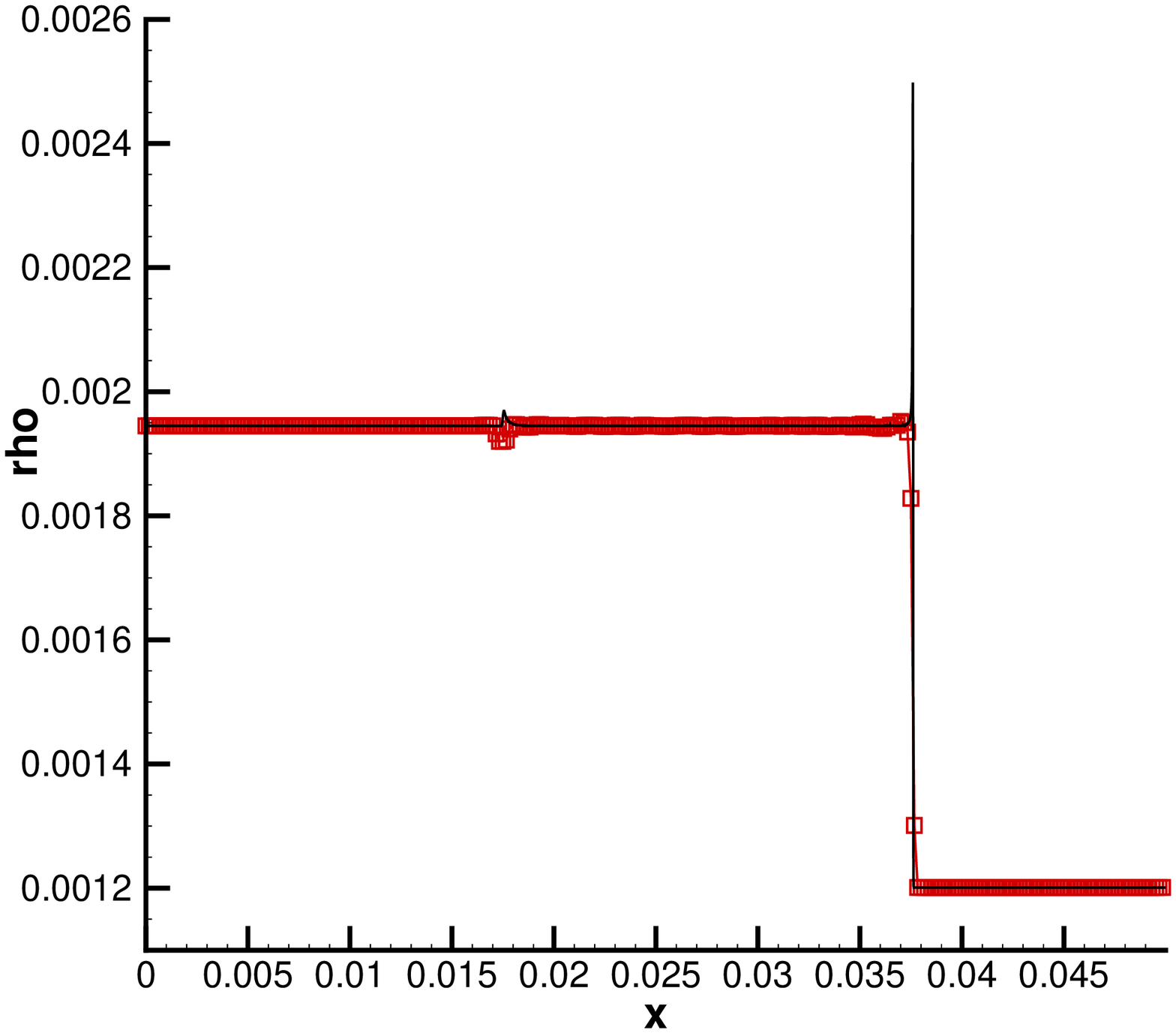}}
	\protect\caption{The same as Fig.~\ref{fig:ex42P} but for density field. \label{fig:ex42rho}}	
\end{figure}

\begin{figure}
	\subfigure[WENO]{\centering\includegraphics[scale=0.3,trim={1.0cm 1.0cm 1.0cm 1.0cm},clip]{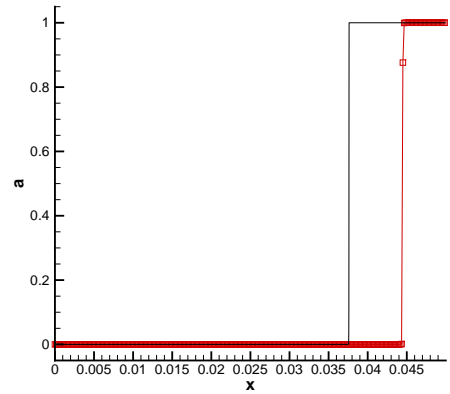}}
	\subfigure[MUSCL-THINC-BVD]{\centering\includegraphics[scale=0.3,trim={1.0cm 1.0cm 1.0cm 1.0cm},clip]{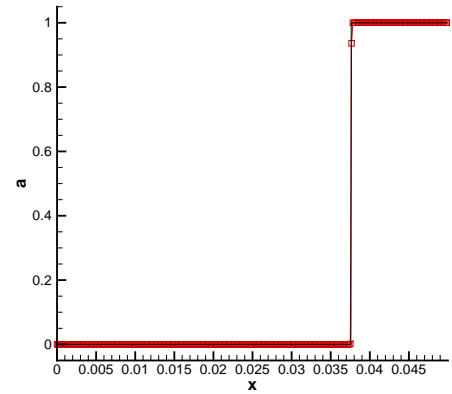}}
	\protect\caption{The same as Fig.~\ref{fig:ex42P} but for mass fraction. \label{fig:ex42a}}	
\end{figure}

The C-J detonation of which the chemical reaction is modeled by the Heaviside formulation is considered in this example. Same as \cite{Random2000,MinMax,Wang2012}, the parameter values are $\frac{1}{\xi}=0.5825\times10^{10}$, $\gamma=1.4$, $q_{0}=0.5196\times10^{10}$ and $T_{ign}=0.1155\times10^{10}$. The totally burnt gas is set on the right side with the initial state as $\rho_{0}=1.201\times10^{-3}$, $u_{0}=0.0$ and $p_{0}=8.321\times10^5$. The totally unburnt gas is set on the left side where $(\rho_{CJ},u_{CJ},p_{CJ},0.0)$ are determined by C-J detonation model. The computational domain is $[0,0.05]$ where discontinuity is set at $x=0.005$. The cell numbers is $N=300$. The computation is evolved until $t=3\times10^{-7}$ with the reaction sub-step $Nr=10$. The exact position of wave will be at $x=0.03764$. We set CFL=0.01 for WENO while CFL=0.1 for MUSCL-THINC-BVD scheme. The results of pressure, temperature, density and mass fraction are plotted in Fig.~\ref{fig:ex42P}-\ref{fig:ex42a}. The correct detonation speed can be captured by proposed MUSCL-THINC-BVD scheme with larger CFL number. However, as reported in \cite{Wang2012} wrong results are produced by standard WENO scheme no matter how small the time step is since the stiffness problem is due to spatial rather than the temporal errors.  

\subsection{A strong detonation}
\begin{figure}
	\subfigure[WENO]{\centering\includegraphics[scale=0.3,trim={1.0cm 1.0cm 1.0cm 1.0cm},clip]{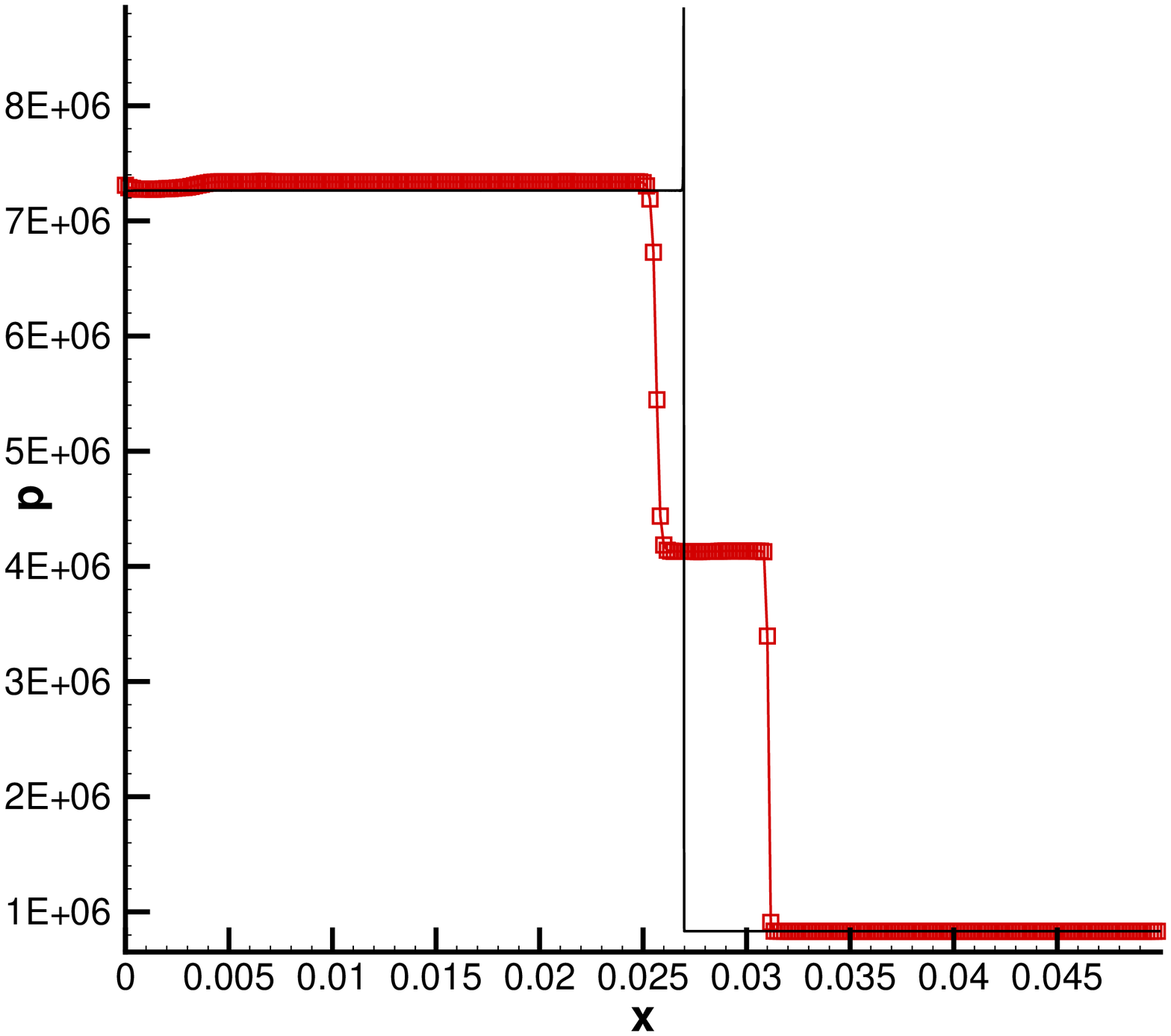}}
	\subfigure[MUSCL-THINC-BVD]{\centering\includegraphics[scale=0.3,trim={1.0cm 1.0cm 1.0cm 1.0cm},clip]{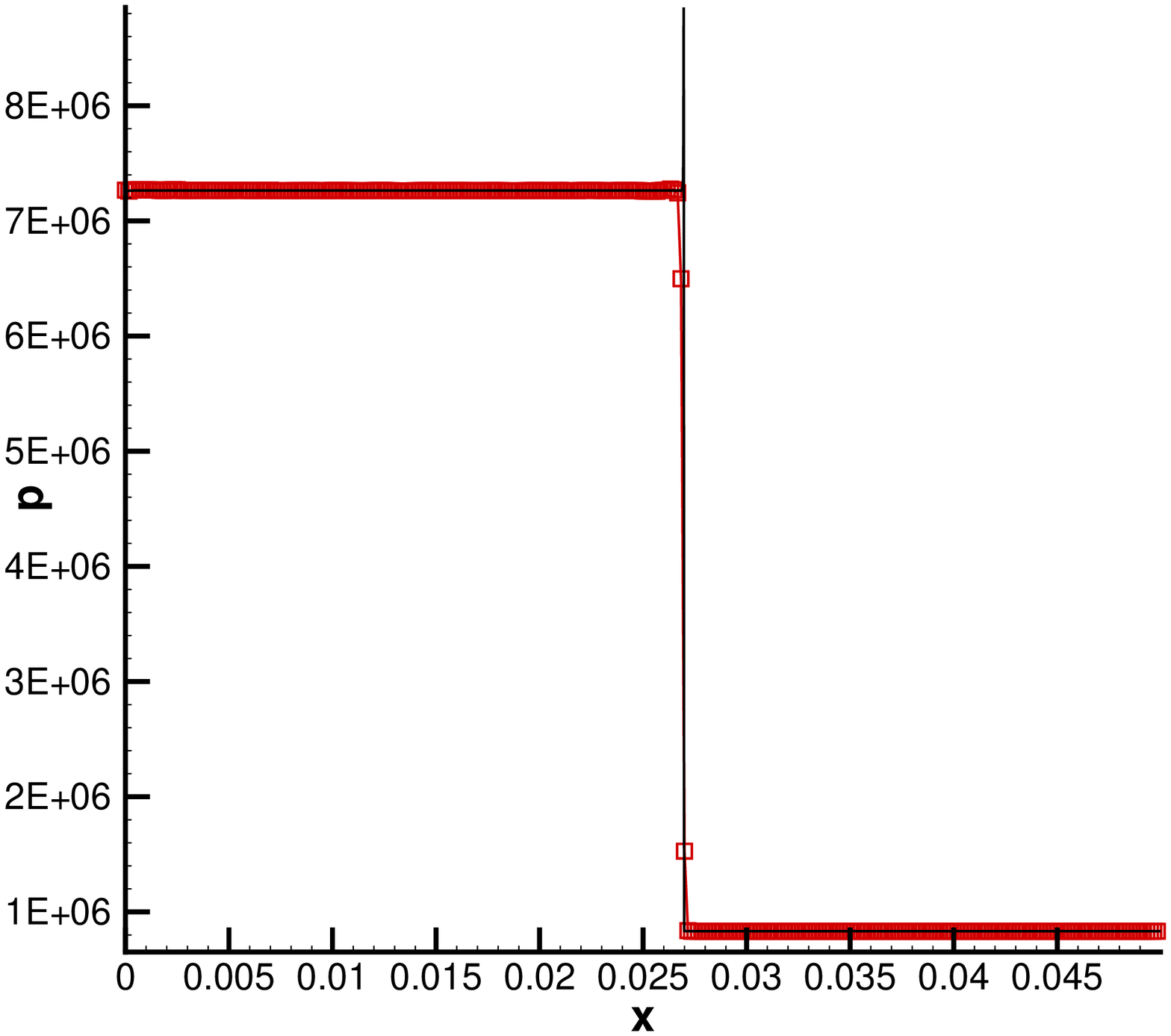}}
	\protect\caption{Numerical results of pressure field for a strong detonation wave. Reference solutions are represented by black solid lines while numerical solutions are represented by red lines with symbols. Comparisons are made between the WENO and MUSCL-THINC-BVD scheme. \label{fig:ex44P}}	
\end{figure}

\begin{figure}
	\subfigure[WENO]{\centering\includegraphics[scale=0.3,trim={1.0cm 1.0cm 1.0cm 1.0cm},clip]{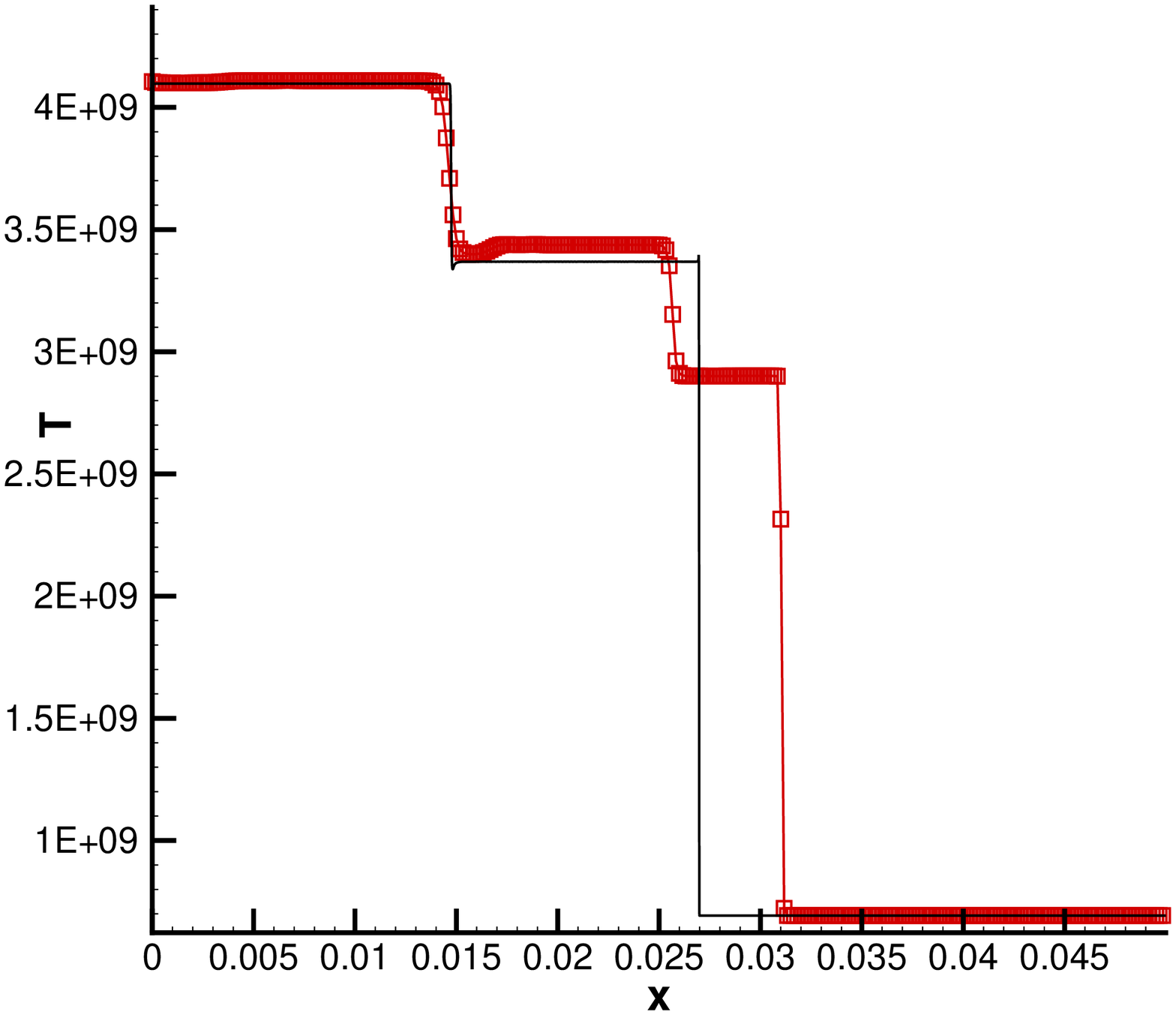}}
	\subfigure[MUSCL-THINC-BVD]{\centering\includegraphics[scale=0.3,trim={1.0cm 1.0cm 1.0cm 1.0cm},clip]{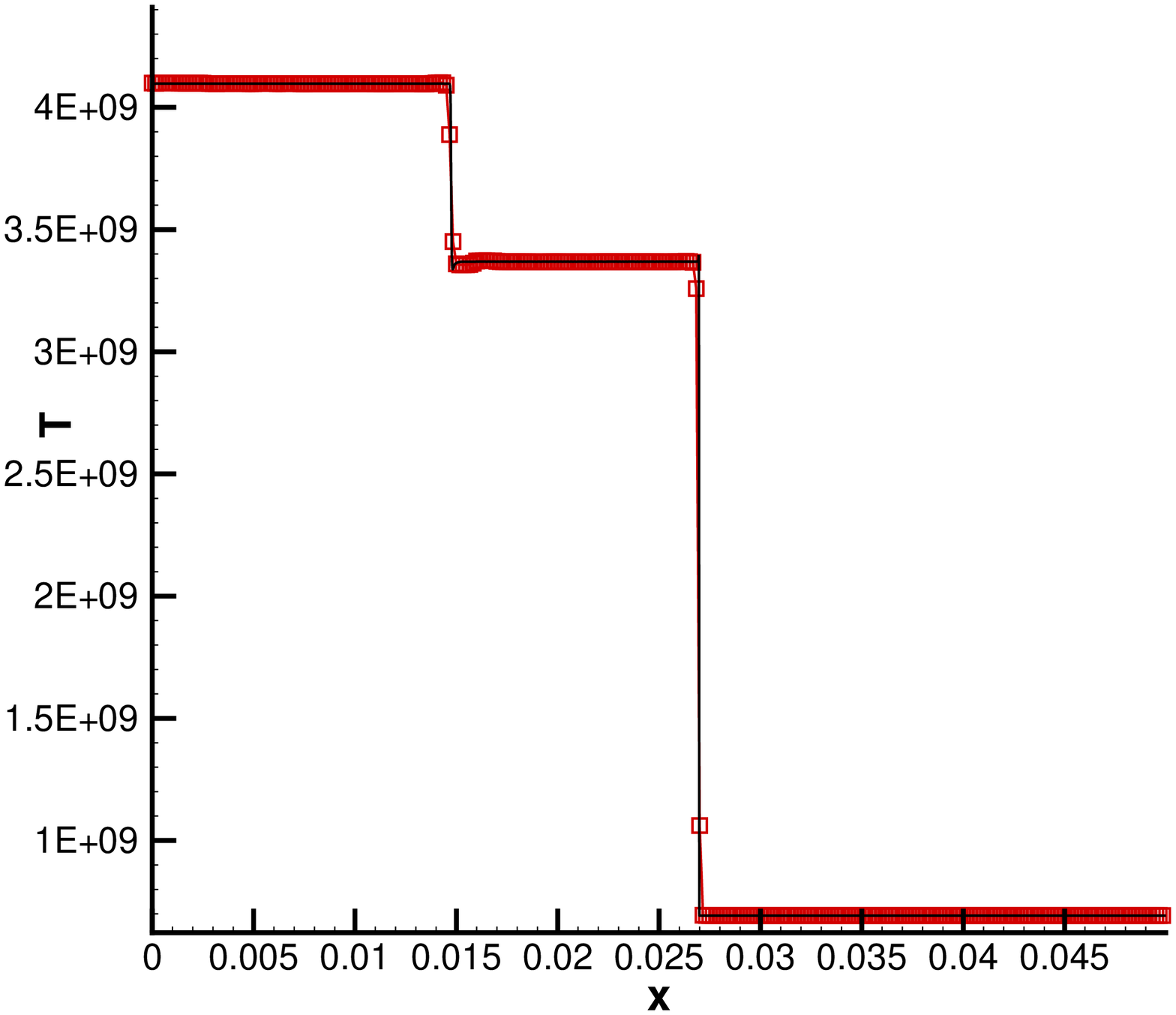}}
	\protect\caption{The same as Fig.~\ref{fig:ex44P} but for temperature field. \label{fig:ex44T}}	
\end{figure}

\begin{figure}
	\subfigure[WENO]{\centering\includegraphics[scale=0.3,trim={0.6cm 1.0cm 1.0cm 1.0cm},clip]{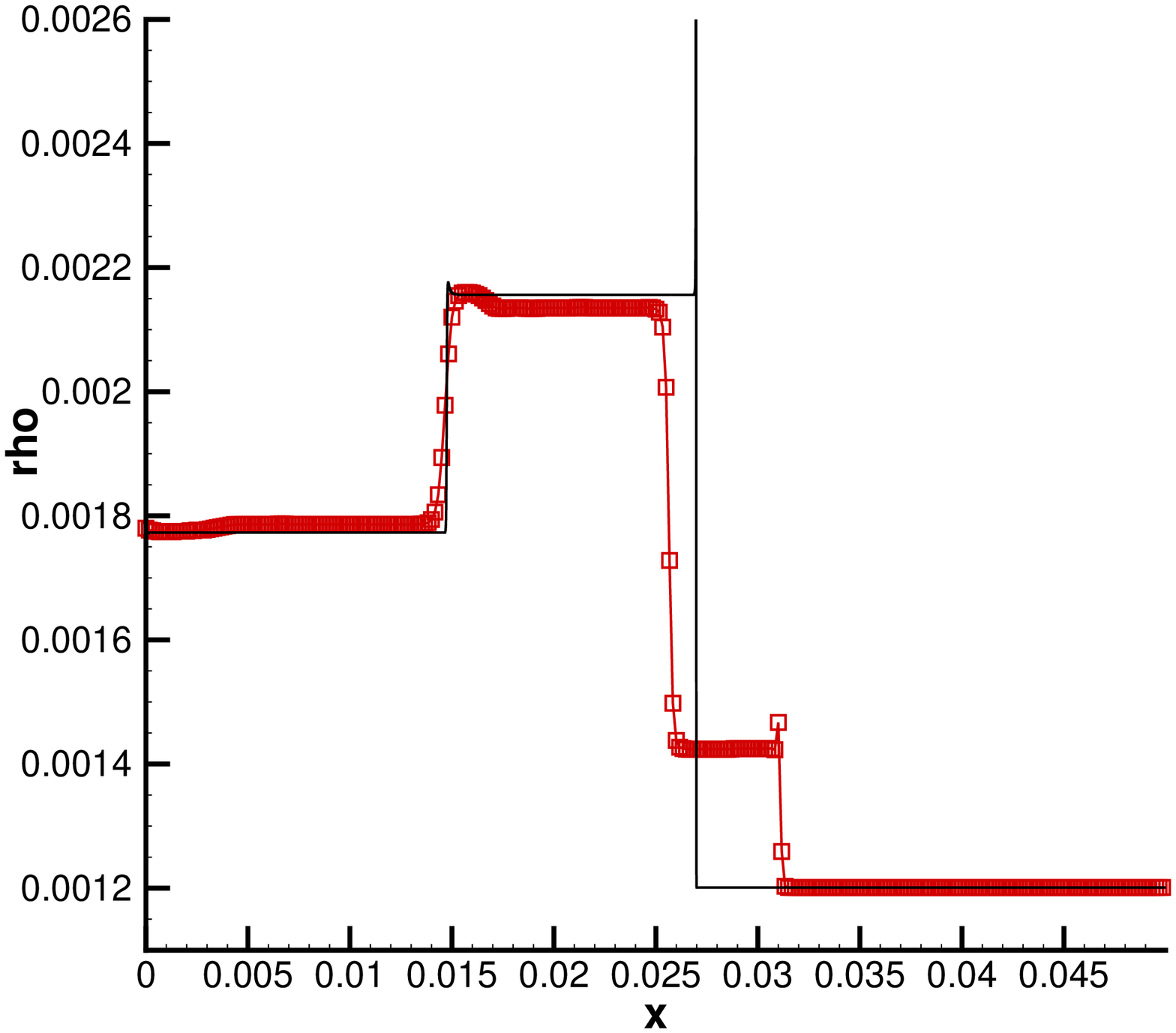}}
	\subfigure[MUSCL-THINC-BVD]{\centering\includegraphics[scale=0.3,trim={0.6cm 1.0cm 1.0cm 1.0cm},clip]{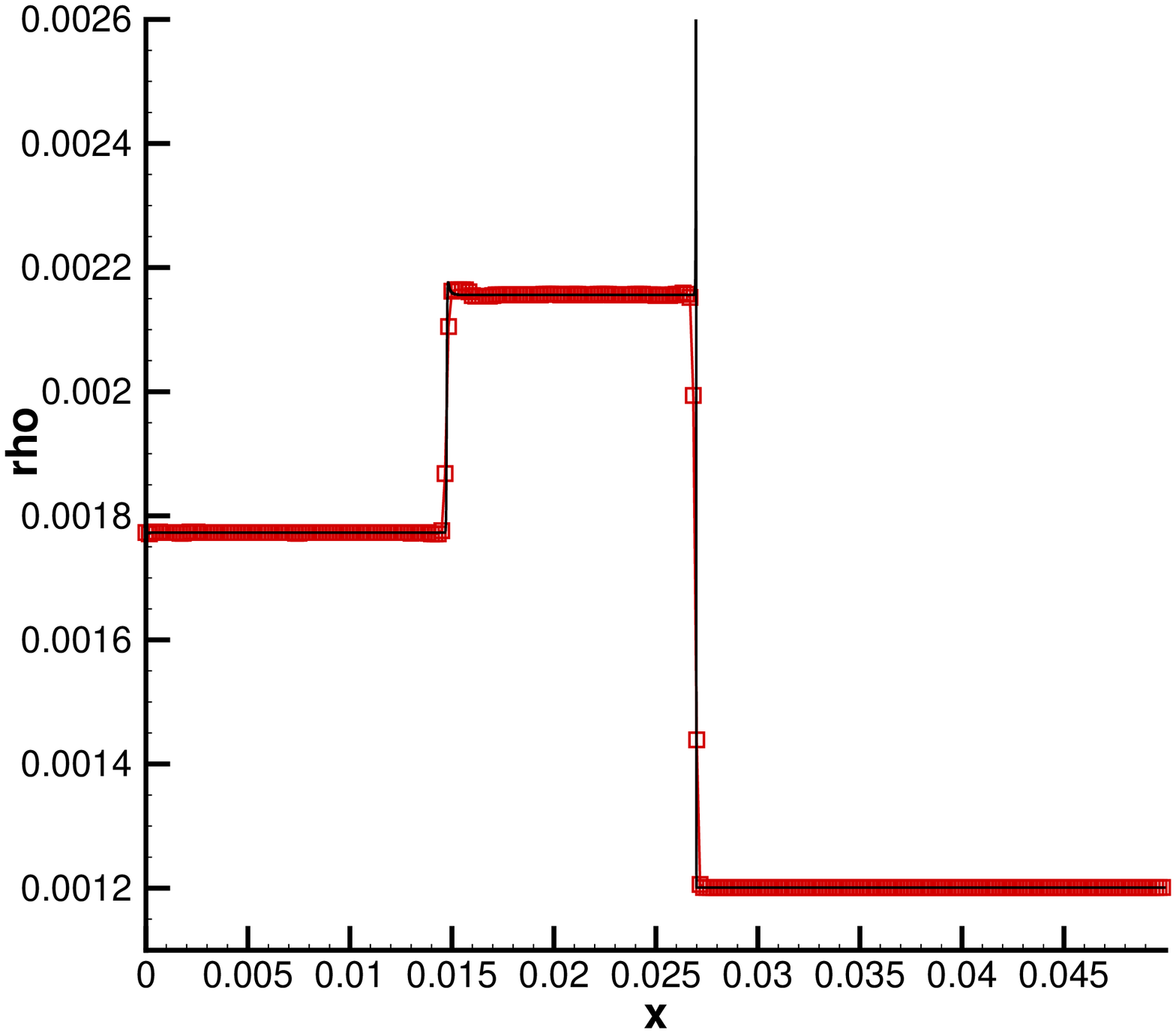}}
	\protect\caption{The same as Fig.~\ref{fig:ex44P} but for density field. \label{fig:ex44rho}}	
\end{figure}

\begin{figure}
	\subfigure[WENO]{\centering\includegraphics[scale=0.3,trim={1.0cm 1.0cm 1.0cm 1.0cm},clip]{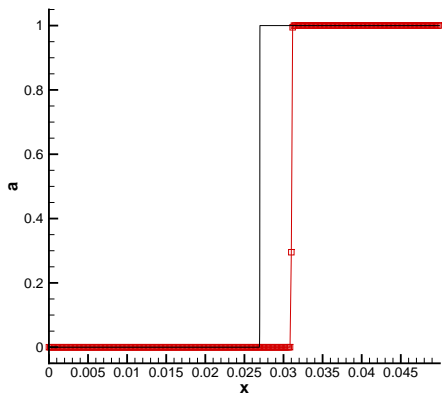}}
	\subfigure[MUSCL-THINC-BVD]{\centering\includegraphics[scale=0.3,trim={1.0cm 1.0cm 1.0cm 1.0cm},clip]{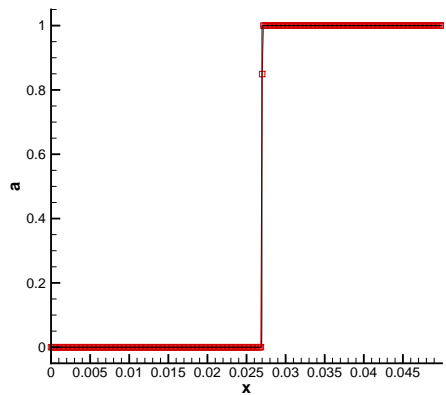}}
	\protect\caption{The same as Fig.~\ref{fig:ex44P} but for mass fraction. \label{fig:ex44a}}	
\end{figure}

Taken from \cite{Random2000}, simulation of a strong detonation is considered here. All parameters are set as the same as the previous example except for pressure at the left side, which is set larger than the value calculated by C-J detonation model. Thus for the initial state at the left side, we have $\rho_{l}=\rho_{CJ}$, $u_{l}=u_{CJ}$ and $p_{l}=8.27\times10^{6}>p_{CJ}$. Different from the profile of C-J detonation, the solution will consist of a right-moving strong detonation wave, a right-moving discontinuity and a left-moving rarefaction wave. The solutions are run to time $t=2\times10^{-7}$ with CFL=0.02 and cell numbers $N=300$. The distributions of pressure, density, temperature and mass fraction field are shown in Fig.~\ref{fig:ex44P}-\ref{fig:ex44a}. It is obvious that WENO scheme produces a spurious weak detonation where the location of the detonation front is essentially wrong. However, MUSCL-THINC-BVD resolves the correct right-moving strong detonation wave. Moreover, regarding to the right-moving discontinuity, it can be solved in only two cells by MUSCL-THINC-BVD scheme while it is solved in more than five points by WENO scheme.

\subsection{Interaction between a detonation wave and an oscillatory profile}
\begin{figure}
	\subfigure[WENO]{\centering\includegraphics[scale=0.3,trim={1.0cm 1.0cm 1.0cm 1.0cm},clip]{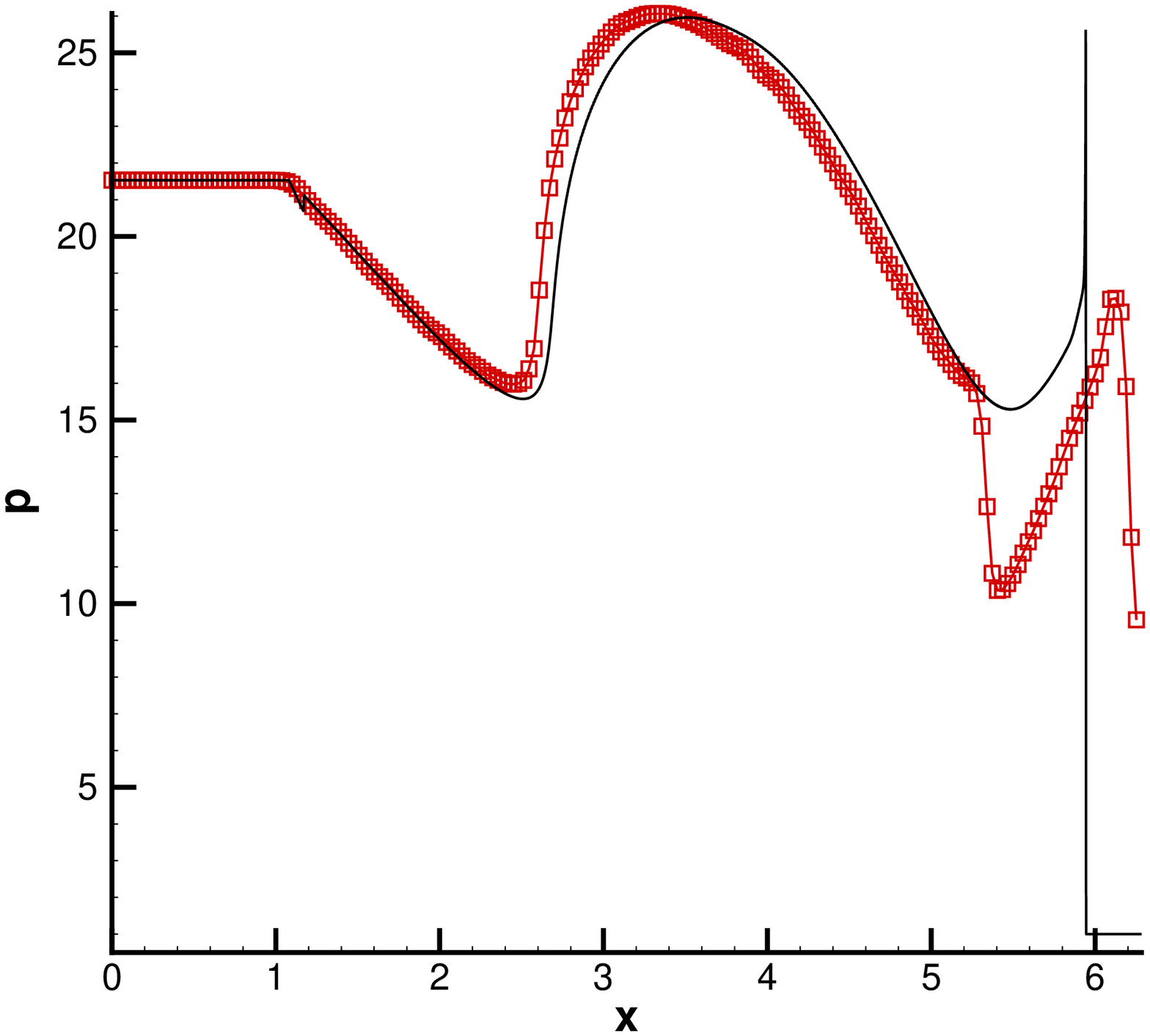}}
	\subfigure[MUSCL-THINC-BVD]{\centering\includegraphics[scale=0.3,trim={1.0cm 1.0cm 1.0cm 1.0cm},clip]{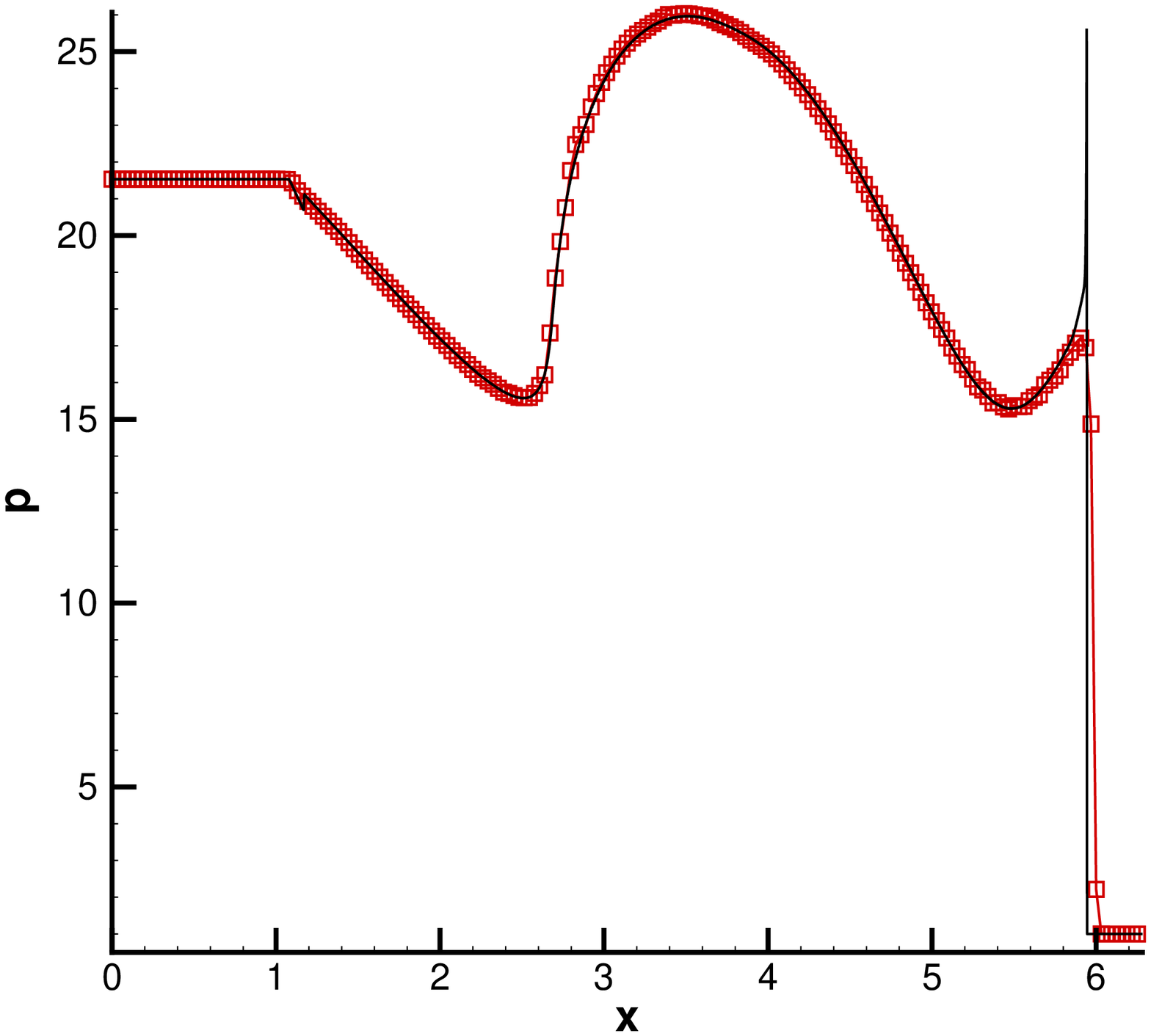}}
	\protect\caption{Numerical results of pressure field for interaction between a detonation wave and an oscillatory profile problem. Reference solutions are represented by black solid lines while numerical solutions are represented by red lines with symbols. Comparisons are made between the WENO and MUSCL-THINC-BVD scheme. \label{fig:ex45P}}	
\end{figure}

\begin{figure}
	\subfigure[WENO]{\centering\includegraphics[scale=0.3,trim={1.0cm 1.0cm 1.0cm 1.0cm},clip]{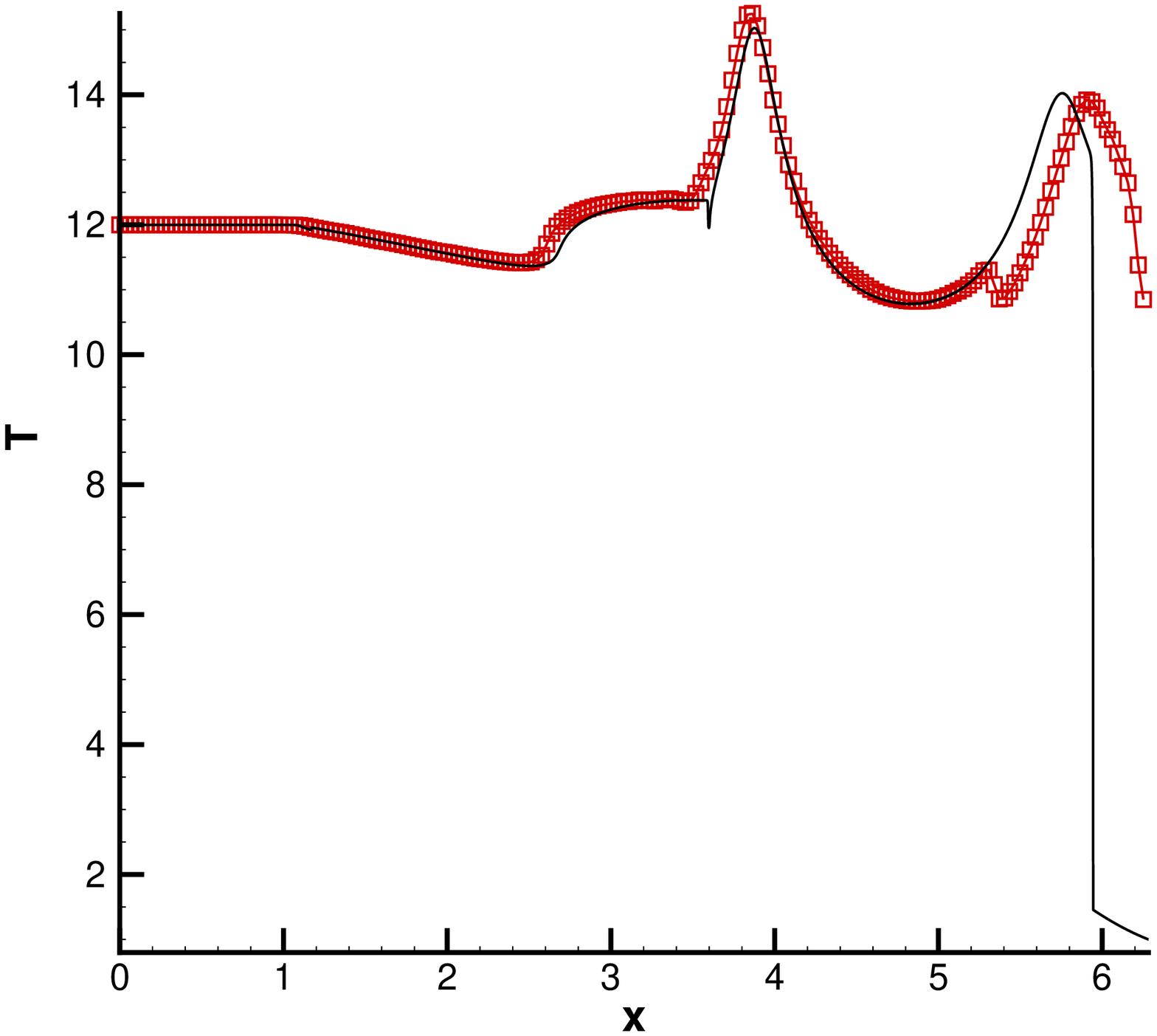}}
	\subfigure[MUSCL-THINC-BVD]{\centering\includegraphics[scale=0.3,trim={1.0cm 1.0cm 1.0cm 1.0cm},clip]{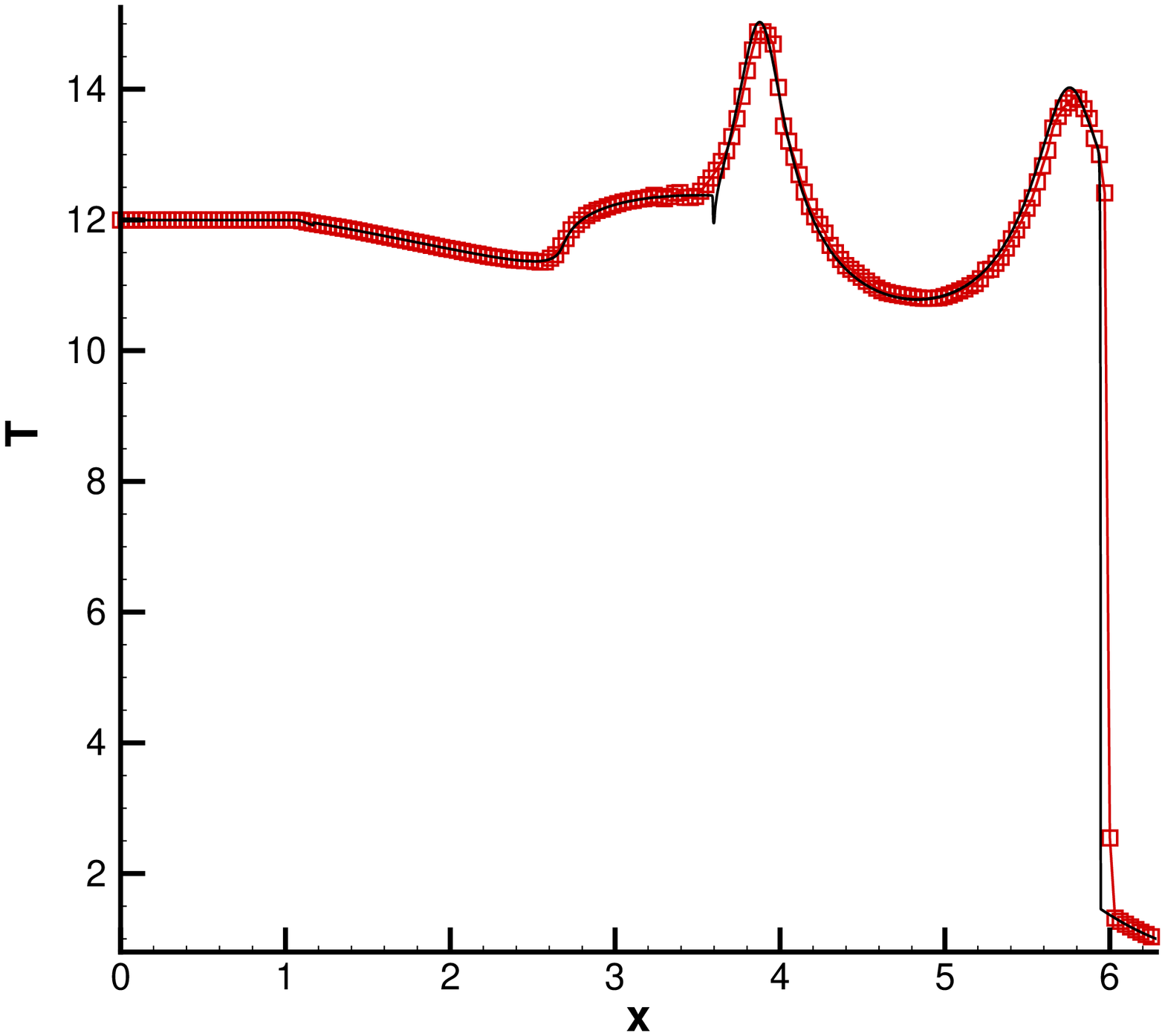}}
	\protect\caption{The same as Fig.~\ref{fig:ex45P} but for temperature field. \label{fig:ex45T}}	
\end{figure}

\begin{figure}
	\subfigure[WENO]{\centering\includegraphics[scale=0.3,trim={0.6cm 1.0cm 1.0cm 1.0cm},clip]{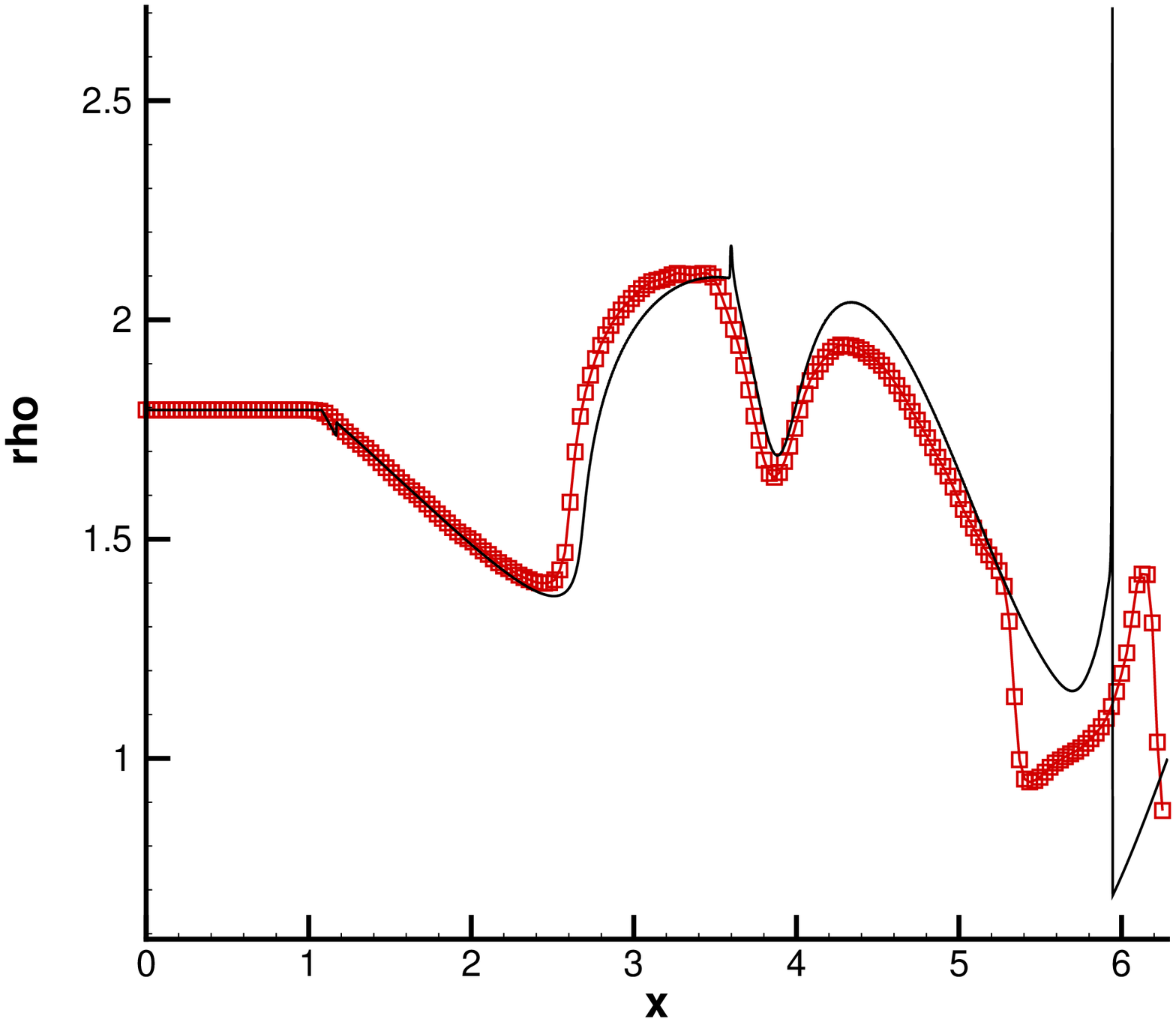}}
	\subfigure[MUSCL-THINC-BVD]{\centering\includegraphics[scale=0.3,trim={0.6cm 1.0cm 1.0cm 1.0cm},clip]{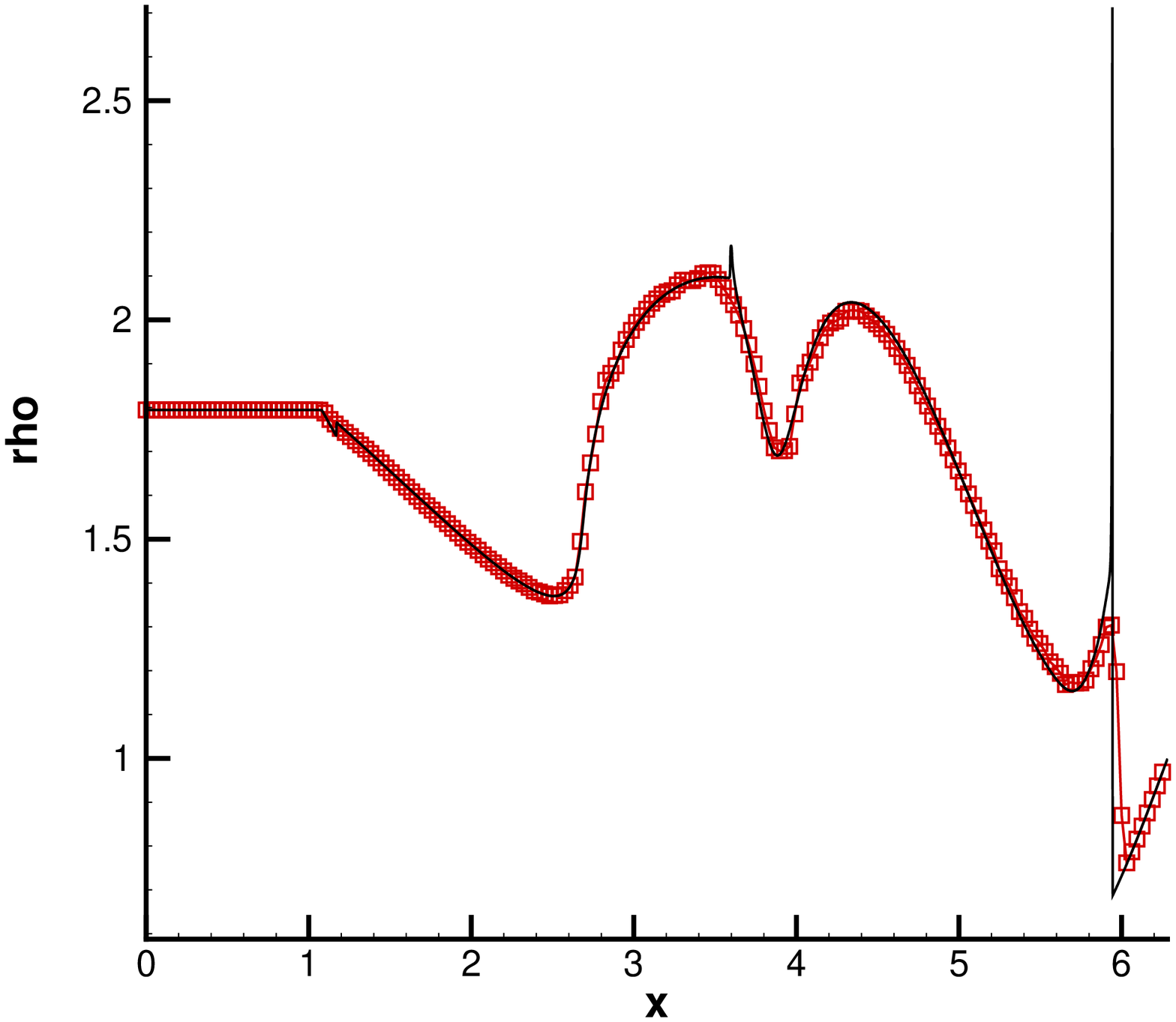}}
	\protect\caption{The same as Fig.~\ref{fig:ex45P} but for density field. \label{fig:ex45rho}}	
\end{figure}

\begin{figure}
	\subfigure[WENO]{\centering\includegraphics[scale=0.3,trim={1.0cm 1.0cm 1.0cm 1.0cm},clip]{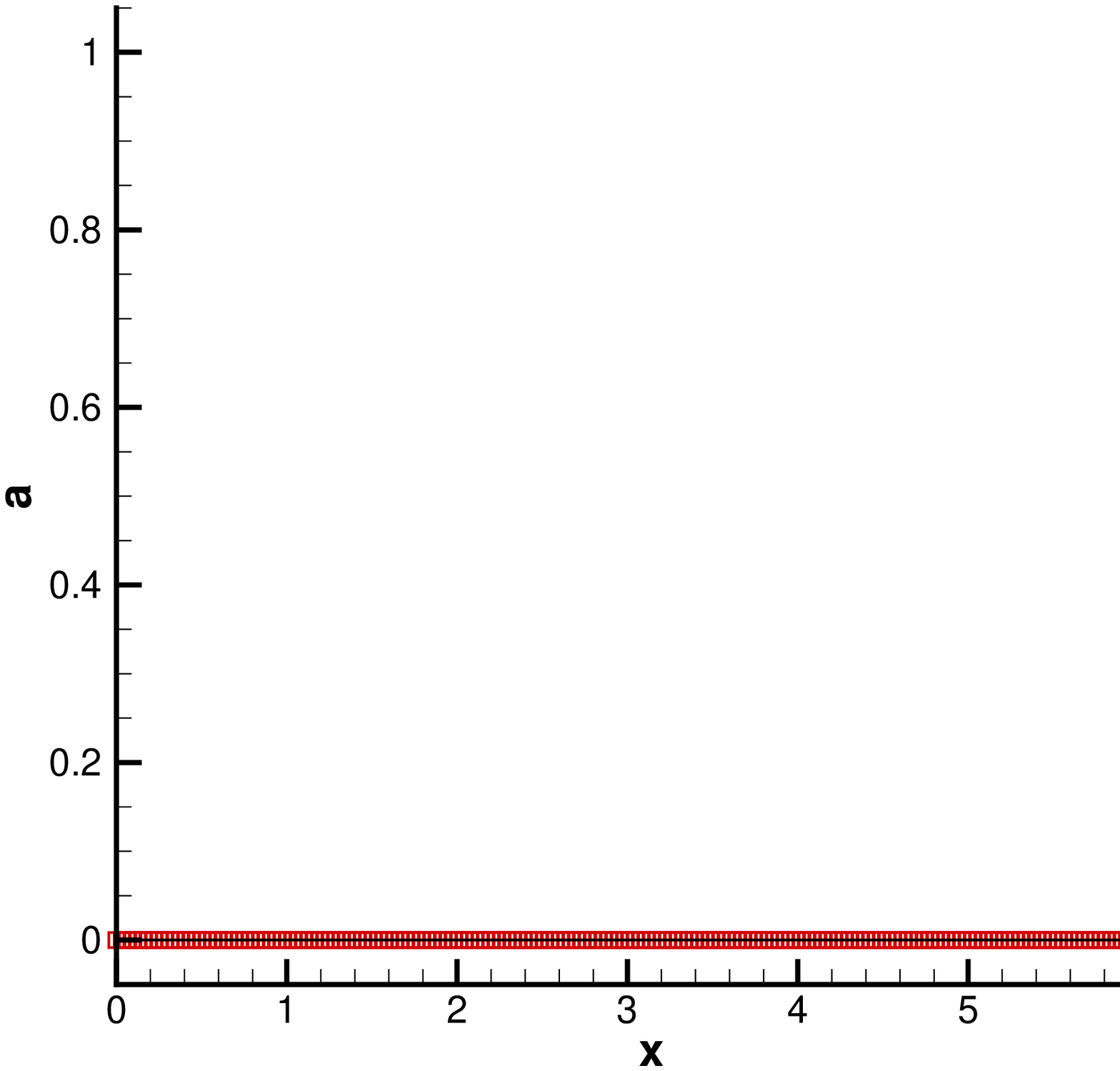}}
	\subfigure[MUSCL-THINC-BVD]{\centering\includegraphics[scale=0.3,trim={1.0cm 1.0cm 1.0cm 1.0cm},clip]{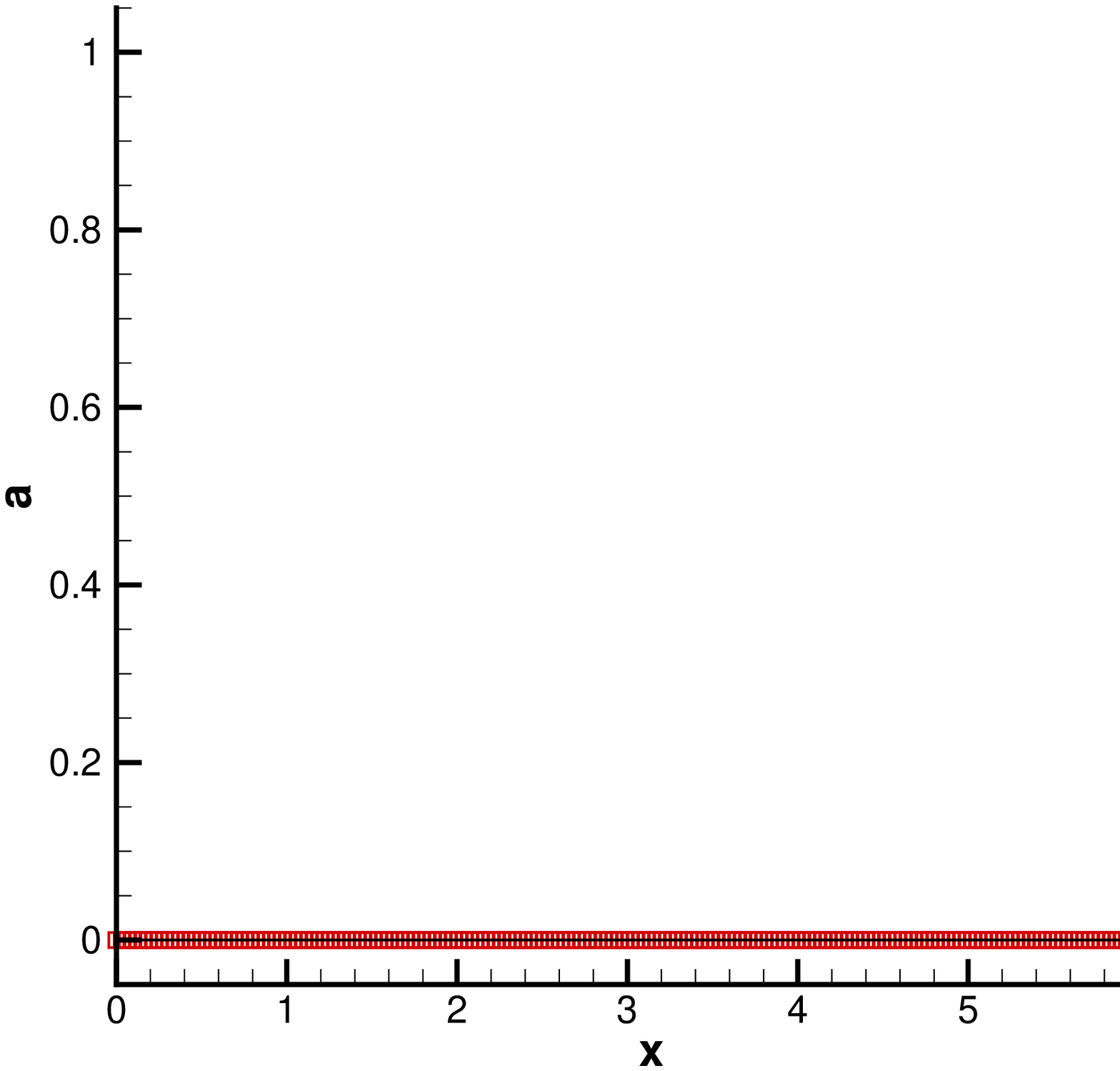}}
	\protect\caption{The same as Fig.~\ref{fig:ex45P} but for mass fraction. \label{fig:ex45a}}	
\end{figure}

Interaction between a detonation wave and an oscillatory profile is considered here to investigate the behavior of the proposed scheme on different field distribution. The same problem has been simulated in \cite{Random01, Wang2012}. The parameters are taken as $\gamma=1.2$, $q_{0}=50$, $\frac{1}{\xi}=1000$ and $T_{ign}=3$. The computational domain is set as $[0,2\pi]$ divided by the mesh numbers $N=200$. The initial state is 
\begin{equation}
(\rho,u,p,\alpha)=\left\{
\begin{array}{l}
(1.79463, 3.0151, 21.53134, 0.0)~~~x\leq \frac{\pi}{2}\\
(1.0+0.5sin2x, 0.0, 1.0, 1.0)~~~\mathrm{otherwise}
\end{array}
\right..
\end{equation}
The computation is run to time $t=\frac{pi}{5}$ with CFL=0.1.The numerical results of density, temperature, pressure and mass fraction fields are plotted in Fig.~\ref{fig:ex45P}-\ref{fig:ex45a}. Still MUSCL-THINC-BVD scheme can prevent the occurrence of spurious waves produced by standard shock capturing WENO scheme. Also, MUSCL-THINC-BVD scheme can resolve the complicated flow field produced by the interaction between detonation waves and oscillatory profiles. 

\subsection{2D detonation waves}
Two dimensional detonation waves problem which has also been investigated by the \cite{Random2000,Wang2012,Yee2013} is considered here. A two-dimensional shock tube with $[0,0.025]\times[0,0.005]$ is set as computational domain. Reflective boundary conditions are prescribed to upper and lower boundaries, while zero-gradient boundary condition are imposed to left and right boundaries. The parameters $q_{0}$, $\frac{1}{\xi}$ and $T_{ign}$ in Heaviside chemical reaction model are same as in Subsection \ref{CJ-H}. The initial condition are
\begin{equation}
(\rho,u,v,p,\alpha)=\left\{
\begin{array}{l}
(\rho_{l}, u_{l}, 0, p_{l}, 0)~~~~\text{if}~x\leq \psi(y)\\
(\rho_{r}, u_{r}, 0, p_{r}, 1)~~~\text{if}~x>\psi(y),
\end{array}
\right.
\end{equation}
where
\begin{equation}
\psi(y)=\left\{
\begin{array}{l}
0.004~~~~~~~~~~~~~~~~~~~~~~~~~~\text{if}~|y-0.0025| \geq 0.001\\
0.005-|y-0.0025|~~~\text{if}~|y-0.0025| < 0.001.
\end{array}
\right.
\end{equation}
The right states $(\rho_{r}, u_{r}, 0, p_{r}, 1)$ are the same as in Subsection \ref{CJ-H} and $\rho_{l}=\rho_{CJ}$, $p_{l}=p_{CJ}$ while $u_{l}=8.162\times10^{4}>u_{CJ}$. As stated in \cite{random19}, one important feature of this solution is that a cellular pattern will form after the triple points travel in the transverse direction and reflect back and forth against the upper and lower boundaries.

\begin{figure}
	\subfigure[Reference]{\centering\includegraphics[scale=0.57,trim={0.0cm 1.0cm 1.0cm 1.0cm},clip]{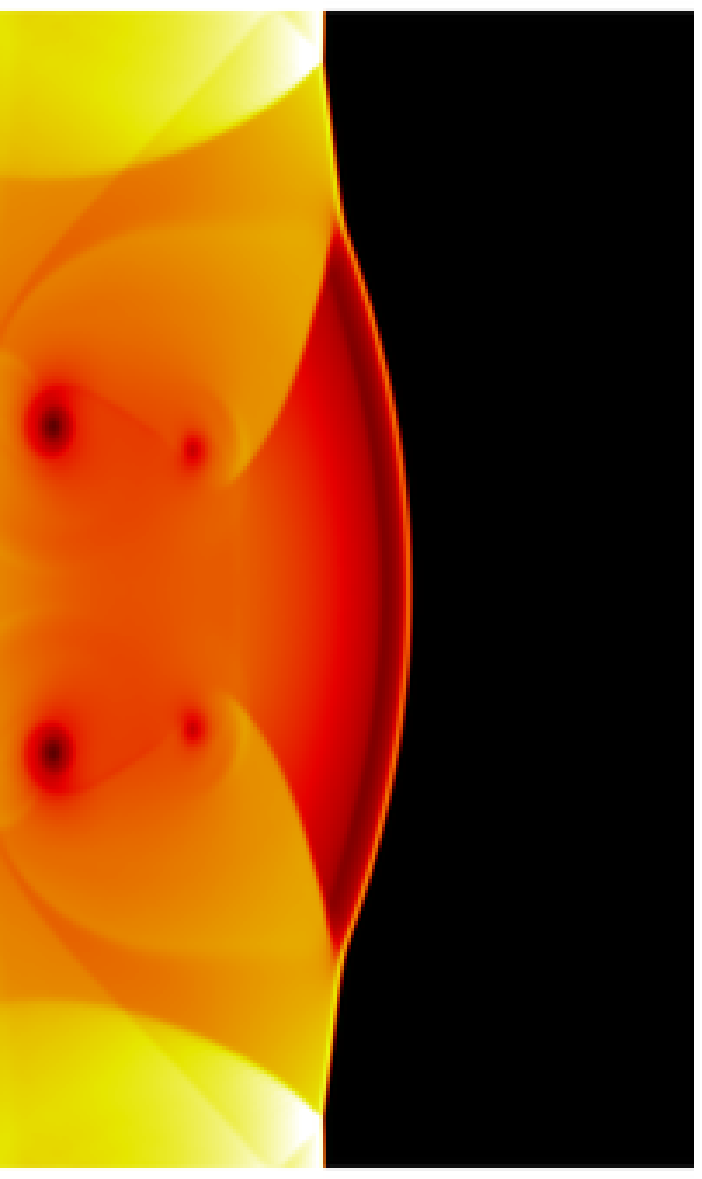}}
	\subfigure[WENO]{\centering\includegraphics[scale=0.54,trim={0.0cm 0.5cm 0.5cm 0.5cm},clip]{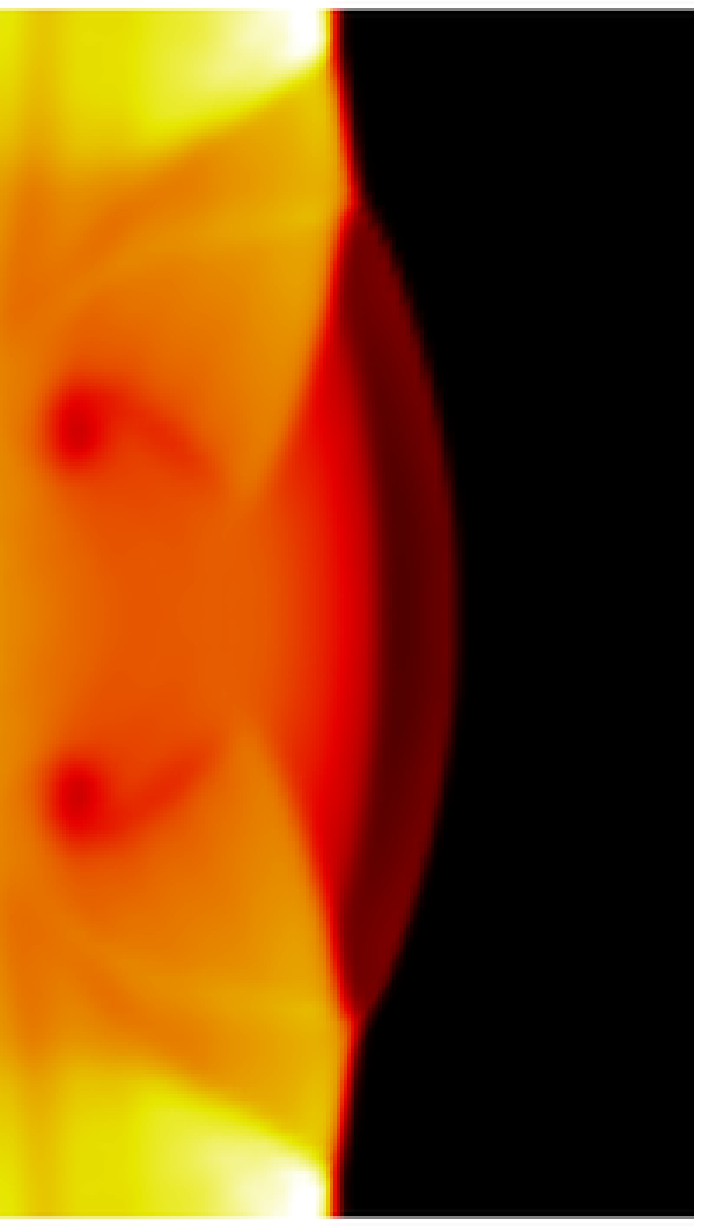}}
	\subfigure[MUSCL-THINC-BVD]{\centering\includegraphics[scale=0.54,trim={0.0cm 0.5cm 0.5cm 0.5cm},clip]{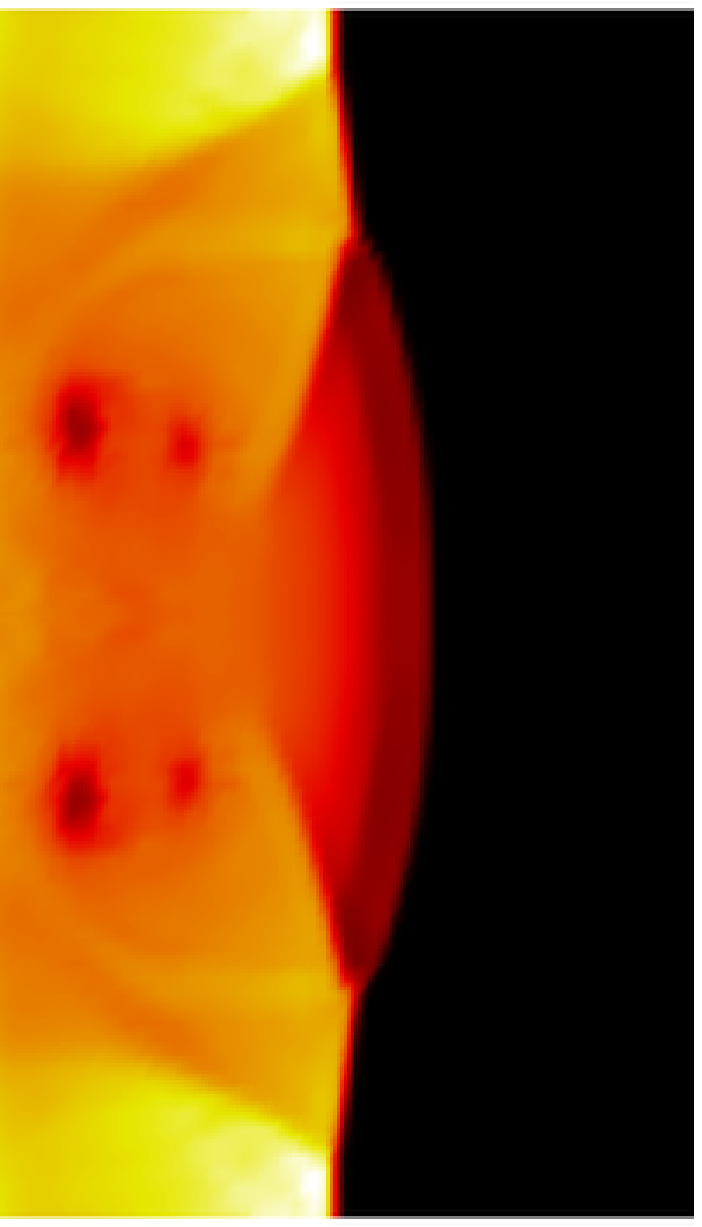}}	
	\protect\caption{Density field of 2D detonation problem at $t=0.3\times10^{-7}$. Comparisons are made among reference solution, WENO scheme with coarse mesh and MUSCL-THINC-BVD with coarse mesh. The profile at the left side is the reference result calculated by WENO scheme with $2000\times 400$ grid cells. The middle profile is the result calculated by WENO scheme with $400\times 80$ grid cells. The right one is calculated by MUSCL-THINC-BVD with $400\times 80$ grid cells.\label{fig:2d9}}	
\end{figure}

\begin{figure}
	\subfigure[Reference]{\centering\includegraphics[scale=0.57,trim={0.0cm 1.0cm 1.0cm 1.0cm},clip]{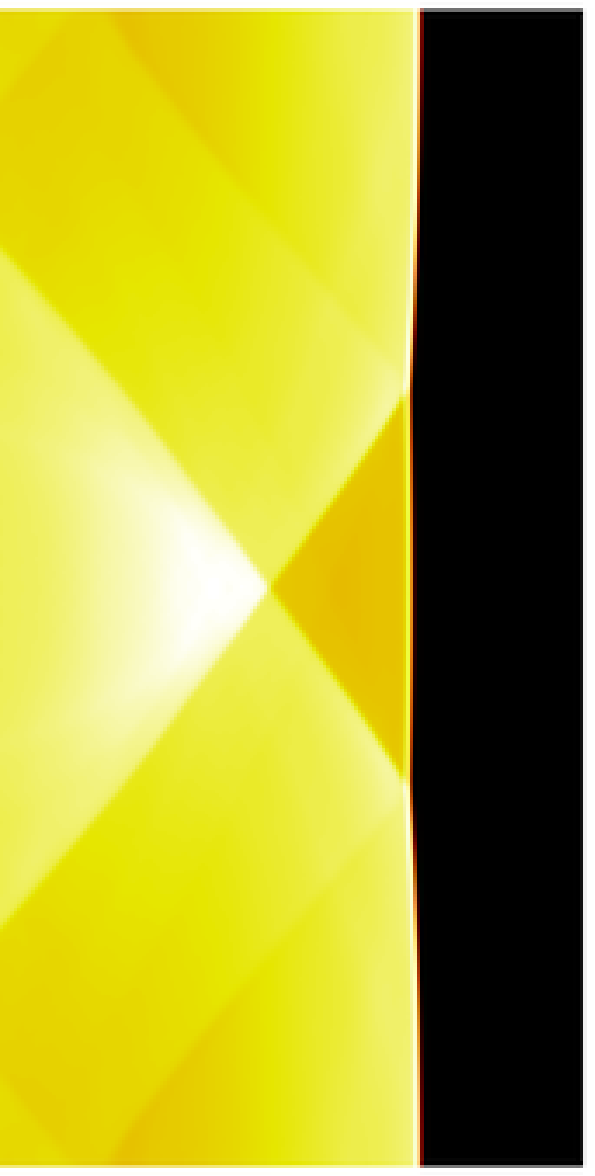}}
	\subfigure[WENO]{\centering\includegraphics[scale=0.54,trim={0.0cm 0.5cm 0.5cm 0.5cm},clip]{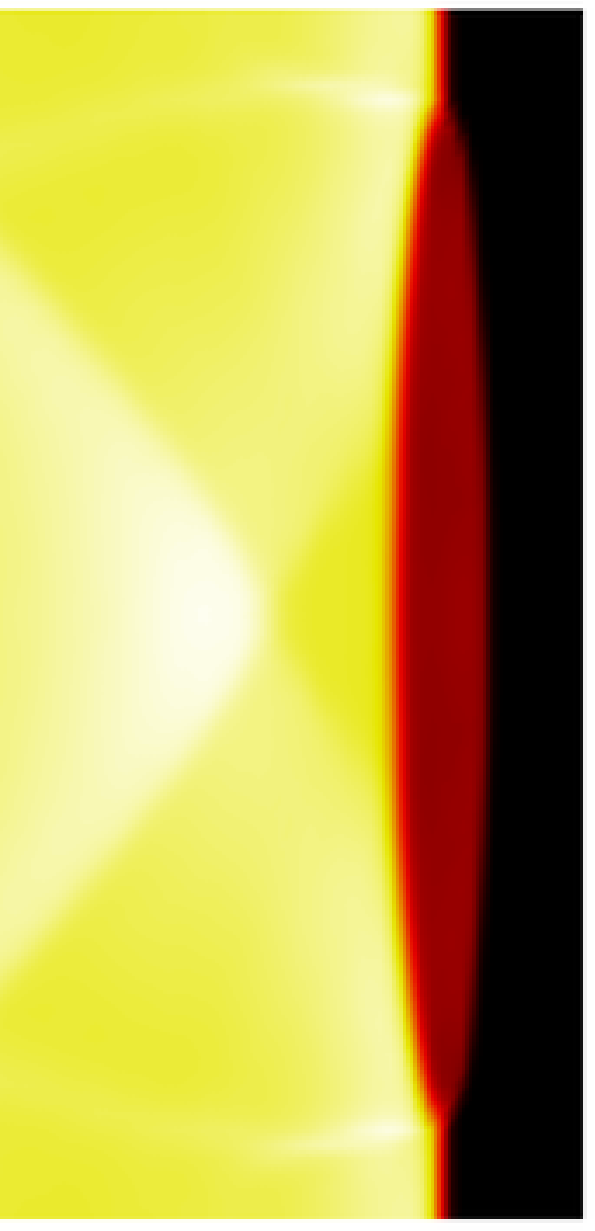}}
	\subfigure[MUSCL-THINC-BVD]{\centering\includegraphics[scale=0.54,trim={0.0cm 0.5cm 0.5cm 0.5cm},clip]{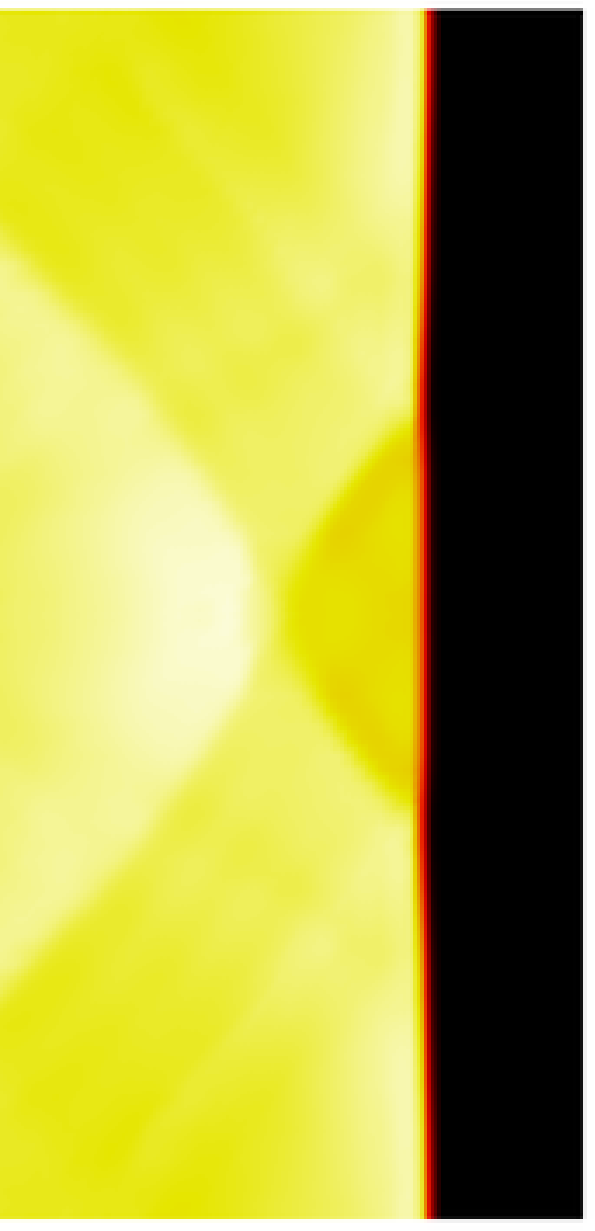}}	
	\protect\caption{The same as Fig.~\ref{fig:2d9} but at $t=1.7\times10^{-7}$ .\label{fig:2d50}}	
\end{figure}

The CFL number is set to 0.1. To make a better comparison the reference solution is calculated by standard 5th order WENO scheme with $2000\times400$ grid cells. At the same time, comparison are made between MUSCL-THINC-BVD and WENO scheme on $400\times80$ cell numbers. The density profiles at different times are presented in Fig.~\ref{fig:2d9}-\ref{fig:2d50} in which the reference solutions are shown in left side while solutions with $400 \times 80$ grid cells by WENO and MUSCL-THINC-BVD are shown at middle and right side respectively. In Fig.~\ref{fig:2d9}, compared with the results computed by MUSCL-THINC-BVD scheme, the reaction zone are smeared by WENO. Moreover, MUSCL-THINC-BVD scheme can resolve the two vortices formed behind the detonation from since as shown in \cite{Xie,dengM,dengAIAA,dengCF} BVD algorithm can improve the resolution quality. In Fig.~\ref{fig:2d50}, spurious waves in front of detonation are produced by WENO scheme while MUSCL-THINC-BVD can capture the sharp detonation front on a coarse mesh.   
\clearpage

\section{Conclusion remarks \label{sec:conclusion}}
A new shock-capturing scheme called MUSCL-THINC-BVD is introduced in present work to solve the stiff detonation waves problems. This method can solve discontinuous solutions with significantly reduced numerical dissipation errors thus prevent wrong propagation speed of discontinuities or spurious waves which are common phenomenon when applying conventional shock-capturing scheme under the insufficient grids resolution. Unlike some existing methods which reply on extra treatment by accepting the diffused discontinuities profiles, the current method resolve the correct position of detonation front by reducing the numerical dissipation errors around discontinuities fundamentally. Thus MUSCL-THINC-BVD scheme is an effective and simple method to simulate stiff detonation problems.

\section*{Acknowledgment}

This work was supported in part by JSPS KAKENHI Grant Numbers 15H03916 and 15J09915. 

\clearpage{}



\begin{thebibliography}{10}
\bibitem{Yee6}Colella, Phillip, Andrew Majda, and Victor Roytburd. "Theoretical and numerical structure for reacting shock waves." SIAM Journal on Scientific and Statistical Computing 7.4 (1986): 1059-1080.	

\bibitem{wang1}Jeltsch, Rolf, and Petra Klingenstein. "Error estimators for the position of discontinuities in hyperbolic conservation laws with source terms which are solved using operator splitting." Computing and Visualization in Science 1.4 (1999): 231-249.

\bibitem{wang7}Bihari, Barna L., and Donald Schwendeman. "Multiresolution schemes for the reactive Euler equations." Journal of Computational Physics 154.1 (1999): 197-230.

\bibitem{wang24}Leveque, Randall J., and Keh-Ming Shyue. "One-dimensional front tracking based on high resolution wave propagation methods." SIAM Journal on Scientific Computing 16.2 (1995): 348-377.

\bibitem{Nguyen}Nguyen, Duc, Frédéric Gibou, and Ronald Fedkiw. "A fully conservative ghost fluid method and stiff detonation waves." 12th Int. Detonation Symposium, San Diego, CA. 2002.

\bibitem{wang14}Engquist, Björn, and Björn Sjögreen. Robust difference approximations of stiff inviscid detonation waves. Department of Mathematics, University of California, Los Angeles, 1991.

\bibitem{Random2000}Bao, Weizhu, and Shi Jin. "The random projection method for hyperbolic conservation laws with stiff reaction terms." Journal of Computational Physics 163.1 (2000): 216-248.

\bibitem{Random01}Bao, Weizhu, and Shi Jin. "The random projection method for stiff detonation capturing." SIAM Journal on Scientific Computing 23.3 (2001): 1000-1026.

\bibitem{Random02}Bao, Weizhu, and Shi Jin. "The random projection method for stiff multispecies detonation capturing." Journal of Computational Physics 178.1 (2002): 37-57.

\bibitem{MinMax}Tosatto, Luca, and Luigi Vigevano. "Numerical solution of under-resolved detonations." Journal of Computational Physics 227.4 (2008): 2317-2343.

\bibitem{Wang2012}Wang, Wei, Chi-Wang Shu, H. C. Yee, and Björn Sjögreen. "High order finite difference methods with subcell resolution for advection equations with stiff source terms." Journal of Computational Physics 231.1 (2012): 190-214.

\bibitem{Harten}Harten, Ami. "ENO schemes with subcell resolution." Journal of Computational Physics 83.1 (1989): 148-184.

\bibitem{Yee2013}Yee, H. C., Dmitry V. Kotov, Wei Wang, and Chi-Wang Shu. "Spurious behavior of shock-capturing methods by the fractional step approach: Problems containing stiff source terms and discontinuities." Journal of Computational Physics 241 (2013): 266-291.

\bibitem{Van_Leer}Van Leer, Bram. ``Towards the ultimate conservative difference scheme. V. A second-order sequel to Godunov's method." Journal of computational Physics 32.1 (1979): 101-136.

\bibitem{xiao_thinc2}Xiao, F., Y. Honma, and T. Kono. ``A simple algebraic interface capturing scheme using hyperbolic tangent function." International Journal for Numerical Methods in Fluids 48.9 (2005): 1023-1040.

\bibitem{Sun}Sun, Ziyao, Satoshi Inaba, and Feng Xiao. ``Boundary Variation Diminishing (BVD) reconstruction: A new approach to improve Godunov schemes." Journal of Computational Physics 322 (2016): 309-325.	

\bibitem{Xie}Xie, Bin, Xi Deng, Ziyao Sun, and Feng Xiao. ``A hybrid pressure?density-based Mach uniform algorithm for 2D Euler equations on unstructured grids by using multi-moment finite volume method." Journal of Computational Physics 335 (2017): 637-663.

\bibitem{dengM}Deng, Xi, Satoshi Inaba, Bin Xie, Keh-Ming Shyue, and Feng Xiao. "Implementation of BVD (boundary variation diminishing) algorithm in simulations of compressible multiphase flows." arXiv preprint arXiv:1704.08041 (2017).

\bibitem{dengAIAA}Deng, Xi, Bin Xie, and Feng Xiao. "Multimoment Finite Volume Solver for Euler Equations on Unstructured Grids." AIAA Journal (2017).

\bibitem{Jiang}Jiang, Guang-Shan, and Chi-Wang Shu. ``Efficient implementation of weighted ENO schemes." Journal of computational physics 126.1 (1996): 202-228.

\bibitem{xiao_thinc}Xiao, Feng, Satoshi Ii, and Chungang Chen. ``Revisit to the THINC scheme: a simple algebraic VOF algorithm." Journal of Computational Physics 230.19 (2011): 7086-7092.
	
\bibitem{wave1}Ketcheson, David I., Matteo Parsani, and Randall J. LeVeque. ``High-order wave propagation algorithms for hyperbolic systems." SIAM Journal on Scientific Computing 35.1 (2013): A351-A377.	

\bibitem{Riemann}Toro, Eleuterio F. Riemann solvers and numerical methods for fluid dynamics: a practical introduction. Springer Science \& Business Media, 2013.

\bibitem{finite}LeVeque, Randall J. Finite volume methods for hyperbolic problems. Vol. 31. Cambridge university press, 2002.

\bibitem{ssp}Gottlieb, Sigal, Chi-Wang Shu, and Eitan Tadmor. ``Strong stability-preserving high-order time discretization methods." SIAM review 43.1 (2001): 89-112.

\bibitem{random5}Chorin, Alexandre Joel. "Random choice methods with applications to reacting gas flow." Journal of computational physics 25.3 (1977): 253-272.

\bibitem{random8}Courant, Richard, and Kurt Otto Friedrichs. Supersonic flow and shock waves. Vol. 21. Springer Science \& Business Media, 1999.

\bibitem{random19}Strehlow, Roger A. "Nature of transverse waves in detonations." Astronautica Acta 14.5 (1969): 539.

\bibitem{dengCF}Deng, Xi, Bin Xie, and Feng Xiao. "A finite volume multi-moment method with boundary variation diminishing principle for Euler equation on three-dimensional hybrid unstructured grids." Computers \& Fluids 153 (2017): 85-101.
\end{thebibliography}
\end{document}